\documentclass[smallextended]{svjour3}       

\smartqed  
\usepackage{graphicx,amssymb}
\usepackage{epstopdf}
\usepackage{algorithm}
\usepackage{algorithmicx}
\usepackage{algpseudocode}
\usepackage[switch,pagewise]{lineno}
\usepackage[usenames]{color}
\usepackage{booktabs}
\usepackage{verbatim}
\usepackage{amsmath}
\usepackage{subfigure}
\usepackage{url}
\usepackage{longtable}
\usepackage{caption}
\usepackage[normalem]{ulem}
\usepackage[T1]{fontenc}
\usepackage{booktabs}
\usepackage{bm}

\begin{document}

\title{A Multiple Search Operator Heuristic for the Max-k-cut Problem}

\author{Fuda Ma         \and
        Jin-Kao Hao*
}

\authorrunning{F. Ma and J.K. Hao} 

\institute{Fuda Ma \at
              LERIA, Universit\'{e} d'Angers, 2 Boulevard Lavoisier, 49045 Angers Cedex 01, France \\
              \email{ma@info.univ-angers.fr}           
           \and
           Jin-Kao Hao* \emph{(Corresponding author)} \at
              LERIA, Universit\'{e} d'Angers, 2 Boulevard Lavoisier, 49045 Angers Cedex 01, France \\
              Institut Universitaire de France, Paris, France\\
							\email{hao@info.univ-angers.fr}
}


\maketitle

\begin{abstract}
The max-k-cut problem is to partition the vertices of a weighted graph $G = (V,E)$ into $k\geq2$  disjoint subsets such that the weight sum of the edges crossing the different subsets is maximized. The problem is referred as the max-cut problem when $k=2$. In this work, we present a multiple operator heuristic (MOH) for the general max-k-cut problem. MOH employs five distinct search operators organized into three search phases to effectively explore the search space. Experiments on two sets of 91 well-known benchmark instances show that the proposed algorithm is highly effective on the max-k-cut problem and improves the current best known results (lower bounds) of most of the tested instances. For the popular special case $k=2$ (i.e., the max-cut problem), MOH also performs remarkably well by discovering 6 improved best known results. We provide additional studies to shed light on the alternative combinations of the employed search operators.
\end{abstract}

\section{Introduction}\label{sec_Intro}
Let $G=(V,E)$ be an undirected graph with vertex set $V = \{1,\ldots,n\}$ and edge set $E \subset V \times V$, each edge $(i,j)\in E$ being associated a weight $w_{ij}\in Z$. Given $k\in [2,n]$, the max-k-cut problem is to partition the vertex set $V$ into $k$ ($k$ is given) disjoint subsets $\{S_{1},S_{2},\ldots,S_{k}\}$,  (i.e., $\displaystyle\mathop{\cup}_{i=1}^k S_i = V, S_{i} \neq \emptyset, S_{i} \cap S_{j} = \emptyset, \forall i \neq j$), such that the sum of weights of the edges from $E$ whose endpoints belong to different subsets is maximized, i.e.,
\begin{equation}\label{MKC_objective}
\begin{split}
 \mathrm{max} \sum_{1 \leq p < q \leq k} \sum_{i \in S_{p}, j \in S_{q}} w_{ij}.\\
\end{split}
\end{equation}
Particularly, when the number of partitions equals 2 (i.e., $k=2$), the problem is referred as the max-cut problem. The max-k-cut is equivalent to the minimum k-partition (MkP) problem which aims to partition the vertex set of a graph into $k$ disjoint subsets so as to minimize the total weight of the edges joining vertices in the same partition \cite{ghaddar2011branch}.

The max-k-cut problem is a classical NP-hard problem in combinatorial optimization and can not be solved exactly in polynomial time \cite{kann1997hardness}. Moreover, when $k=2$, the max-cut problem is one of the Karp's 21 NP-complete problems \cite{karp1972reducibility} which has been the subject of many studies in the literature.

In recent decades, the max-k-cut problem has attracted increasing attention for its applicability to numerous important applications in the area of data mining \cite{ding2001min}, VLSI layout design \cite{barahona1988application,chang1987efficient,chen1983graph,cho1998fast,pinter1984optimal},  frequency planning  \cite{eisenblatter2002semidefinite}, sports team scheduling  \cite{mitchell2003realignment}, and statistical physics \cite{liers2004computing} among others.

Given its theoretical significance and large application potential, a number of solution procedures for solving the max-k-cut problem (or its equivalent MkP) have been reported in the literature. In \cite{ghaddar2011branch}, the authors provide a review of several exact algorithms which are based on branch-and-cut and semidefinite programming approaches. But due to the high computational complexity of the problem, only instances of reduced size (i.e., $|V| < 100$) can be solved by these exact methods in a reasonable computing time.

For large instances, heuristic and metaheuristic methods are commonly used to find "good-enough'' sub-optimal solutions. In particular, for the very popular max-cut problem, many heuristic algorithms have been proposed, including simulated annealing and tabu search \cite{arraiz2009competitive}, breakout local search \cite{benlic2013breakout}, projected gradient approach \cite{burer2001projected}, discrete dynamic convexized method\cite{lin2012discrete},  rank-2 relaxation heuristic \cite{burer2002rank}, variable neighborhood search \cite{festa2002randomized}, greedy heuristics \cite{kahruman2007greedy}, scatter search \cite{marti2009advanced}, global equilibrium search \cite{shylo2012solving} and its parallel version \cite{shyloGlover2015}, memetic search \cite{wu2012memetic,lin2014efficient}, and unconstrained binary quadratic optimization \cite{wang2013probabilistic}. Compared with max-cut, there are much fewer heuristics for the general max-k-cut problem or its equivalent MkP. Among the rare existing studies, we mention the very recent discrete dynamic convexized method of \cite{Zhu2013}, which formulates the max-k-cut problem as an explicit mathematical model and uses an auxiliary function based local search to find satisfactory results.

In this paper, we partially fill the gap by presenting a new and effective heuristic algorithm for the general max-k-cut problem. The main originality of the proposed algorithm is its multi-phased multi-strategy approach which relies on five distinct local search operators for solution transformations. These operators are organized into three different search phases (descent-based improvement, diversified improvement, perturbation) to ensure an effective examination of the search space. The basic idea of our approach is as follows. The descent-based improvement procedure aims to locate a good local optimum from an initiating solution. This is achieved with two dedicated intensification operators. Then the diversified improvement phase discovers promising areas around the obtained local optimum by applying two additional operators. Once an improved solution is found, the search switches back to the descent-based improvement phase to make an intensive exploitation of the regional area. If the search is trapped in a deep local optimum, the perturbation phase applies a random search operator to definitively lead the search to a distant region from which a new round of the three-phased search procedure starts. This process is repeated until a stop condition is met.

We assess the performance of the proposed algorithm on two sets of well-known benchmarks with a total of 91 instances which are commonly used to test max-k-cut and max-cut algorithms in the literature. Computational results show that the proposed algorithm competes very favorably with respect to the existing max-k-cut heuristics, by improving the current best known results on most instances. Moreover, when the algorithm is applied to the very popular max-cut problem with $k=2$, the results yielded by our algorithm remain highly competitive compared with the most effective and dedicated max-cut algorithms. In particular, for 6 (large) instances, our algorithm manages to improve the current best known solutions reported by any existing specific max-cut algorithms of the literature.

The rest of the paper is organized as follows. In Section \ref{sec_Algorithm}, the proposed algorithm is fully presented. Section \ref{sec_Results} provides computational results and comparisons with other state-of-the-art algorithms in the literature. Section \ref{sec_Discussion} is dedicated to a analysis of several essential parts of the proposed algorithm. Concluding remarks are given in Section \ref{sec_Conclusion}.

\section{Multiple search operator heuristic for max-k-cut} \label{sec_Algorithm}
\subsection{General working scheme} \label{subsec_gen-proc}
The proposed multiple operator heuristic algorithm (MOH) for the general max-k-cut problem is described in Algorithm \ref{algo_mkc} whose components are explained in the following subsections. The algorithm explores the search space (Section \ref{subsec_sspace}) by alternately applying five distinct search operators ($O_1$ to $O_5$) to make transitions from the current solution to a neighboring solution (Section \ref{subsec_mov_ops}). Basically, from an initial solution, the algorithm makes, with two operators ($O_1$ and $O_2$), a descent local search to reach a local optimum $I$ (Alg. \ref{algo_mkc}, lines $11-21$, descent-based improvement phase, Section \ref{subsec_descent_search}). Then the algorithm continues to the diversified improvement phase (Alg. \ref{algo_mkc}, lines $30-40$, Section \ref{subsec_div_search}) which applies two other operators ($O_3$ and $O_4$) to locate new promising regions around the local optimum $I$. This second phase ends each time a better solution than the current local optimum $I$ is discovered or when a maximum number of diversified moves $\omega$ is reached. In both cases, the search returns to the descent-based improvement phase with the best solution found as its new starting point. If no improvement is obtained in $\xi$ descent-based improvement and diversified improvement phases, the search is judged to be trapped in a deep local optimum. To escape this deep local optimum and jump to an unexplored region, the search turns into a perturbation-based diversification phase (Alg. \ref{algo_mkc}, lines $42-45$), which uses a random operator ($O_5$) to strongly transform the current solution (Section \ref{subsec_div_pertub}). The perturbed solution serves then as the new starting solution of the next round of the descent-based improvement phase. This process is iterated until the stop criterion is met.

\begin{algorithm}[t]
\begin{scriptsize}
\caption{General procedure for the max-k-cut problem}\label{algo_mkc}
 \begin{algorithmic}[1]
   \State \sf \textbf{Require}: Graph $G=(V,E)$, number of partitions $k$, max number $\omega$ of diversified moves,  max number $\xi$ of consecutive non-improvement rounds of the descent improvement and diversified improvement phases before the perturbation phase, probability $\rho$ for applying operator $O_3$, $\gamma$ the perturbation strength.
   \State \textbf{Ensure}: the best solution $I_{best}$ found so far

   \State $I \leftarrow$ Generate\_initial\_solution($V,k$) \Comment{  $I$ is a partition of  $V$ into $k$ subsets }
   \State $I_{best}\leftarrow I$  \Comment{ $I_{best}$ Records the best solution found so far }
    \State $f_{lo} \leftarrow f(I)$  \Comment{ {$f_{lo}$ Records the objective value of the latest local optimum reached by  $O_1 \cup O_2$ } }
   \State $f_{best}\leftarrow f(I)$  \Comment{ $f_{best}$ Records the best objective value found so far }
   \State $c_{non\_impv} \leftarrow 0$  \Comment{Counter of consecutive non-improvement rounds of descent and diversified search }
   \State $Iter \leftarrow 0$  \Comment{ Iteration counter }

   \While{stop condition not satisfied}

        \State{/* lines \ref{begin_desent} to \ref{end_desent}: Descent-based improvement phase by applying $O_1$ and $O_2$, see Section \ref{subsec_mov_ops}*/}
	\Repeat  \label{begin_desent}
        \While{$f(I\oplus O_1) > f(I)$}  \Comment{Descent Phase by applying operator $O_1$}
            \State $I \leftarrow I\oplus O_1$ \Comment{ Perform the move defined by $O_1$}
            \State Update $\Delta$  \Comment {$\Delta$ is the bucket structure recording move gains for vertices, see Section \ref{bucket_sort}}
            \State $Iter \leftarrow Iter+1$
        \EndWhile

        \If{$f(I\oplus O_2) > f(I)$} \Comment{Descent Phase by applying operator $O_2$}
            \State $I \leftarrow I\oplus O_2$
            \State Update $\Delta$; $Iter \leftarrow Iter+1$;
       	\EndIf
	\Until{$I$ can not be improved by operator $O_1$ and $O_2$} \label{end_desent}

	   \State $f_{lo} \leftarrow f(I)$
       \If{$f(I)>f_{best}$}
			\State $f_{best} \leftarrow f(I)$; $I_{best} \leftarrow I$\Comment{Update the best solution found so far}
            \State $c_{non\_impv} \leftarrow 0$ \Comment{Reset counter $c_{non\_impv}$}
        \Else
            \State $c_{non\_impv} \leftarrow c_{non\_impv}+1$
		\EndIf	
        \State{/* lines \ref{begin_div} to \ref{end_div}: Diversified improv. phase by applying $O_3$ and $O_4$ at most $\omega$ times, see Section \ref{subsec_mov_ops} */}
        \State $c_{div}\leftarrow 0$ \Comment{Counter $c_{div}$ records number of diversified moves} \label{begin_div}
	    \Repeat
            \If{$Random(0,1) < \rho$} \Comment{Random(0,1) returns a random real number between 0 to 1}
                \State  $I \leftarrow I \oplus O_3$

            \Else
                \State  $I \leftarrow I \oplus O_4$
            \EndIf

            \State Update $H$ ($H,Iter,\lambda$) \Comment{Update tabu list $H$ where $\lambda$ is the tabu tenure, see Section \ref{subsec_mov_ops}}
            \State Update $\Delta$ 
						\State $Iter \leftarrow Iter+1$; $c_{div} \leftarrow c_{div}+1$
            \Until{$c_{div}>\omega$ or $f(I)>f_{lo}$} \label{end_div}

        \State{/* Perturbation phase by applying $O_5$ if $f_{best}$ not improved for $\xi$ rounds of phases 1-2, see Sect. \ref{subsec_div_pertub} */}
        \If{$c_{non\_impv}>\xi$}
        \State $I \leftarrow I \oplus O_5$ \Comment{Apply random perturbation $\gamma$ times, see Section \ref{subsec_div_pertub}}
        \State $c_{non\_impv}\leftarrow 0$ 	
        \EndIf		
  \EndWhile	

\end{algorithmic}
\end{scriptsize}
\end{algorithm}

\subsection{Search space and evaluation solution} \label{subsec_sspace}
Recall that the goal of max-k-cut is to partition the vertex set $V$ into $k$ subsets such that the sum of weights of the edges between the different subsets is maximized.  As such, we define the search space $\Omega$ explored by our algorithm as the set of all possible partitions of $V$ into $k$ disjoint subsets, $\Omega = \{\{S_{1},S_{2},\ldots,S_{k}\}: \displaystyle \mathop{\cup}_{i=1}^k S_i = V, S_{i} \cap S_{j} = \emptyset, S_{i}\subset V, \forall i \neq j$\}, where each candidate solution is called a $k$-cut.

For a given partition or $k$-cut  $I=\{S_{1},S_{2},\ldots,S_{k}\} \in \Omega$, its objective value $f(I)$ is the sum of weights of the edges connecting two different subsets:
\begin{equation}\label{Fitness}
\begin{split}
f(I) =  \sum_{1 \leq p < q \leq k} \sum_{i \in S_{p}, j \in S_{q}} w_{ij}.\\
\end{split}
\end{equation}
Then, for two candidate solutions $I' \in \Omega$ and $I'' \in \Omega$, $I'$ is better than $I''$ if and only if $f(I')>f(I'')$.  The goal of our algorithm is to find a solution $I_{best} \in \Omega$ with $f(I_{best})$ as large as possible.

\subsection{Initial solution}\label{Initial solution}
The MOH algorithm needs an initial solution to start its search. Generally, the initial solution can be provided by any means. In our case, we adopt a randomized two step procedure. First, from $k$ empty subsets $S_i = \emptyset, \forall i\in \{1,\dots,k\}$, we assign each vertex $v \in V$ to a random subset $S_i \in \{S_{1},S_{2},\ldots,S_{k}\}$. Then if some subsets are still empty, we repetitively move a vertex from its current subset to an empty subset until no empty subset exists.

\subsection{Move operations and search operators}\label{subsec_mov_ops}
Our MOH algorithm iteratively transforms the incumbent solution to a neighboring solution by applying some \textit{move} operations. Typically, a move operation (or simply a move) changes slightly the solution, e.g., by transferring a vertex to a new subset. Formally, let $I$ be the incumbent solution and let $mv$ be a move, we use $I' \leftarrow I \oplus mv$ to denote the neighboring solution $I'$ obtained by applying $mv$ to $I$.

Associated to a move operation $mv$, we define the notion of \textit{move gain} $\Delta_{mv}$, which indicates the objective change between the incumbent solution and the neighboring solution obtained by applying the move, i.e.,
\begin{equation}\label{form-mgain}
\Delta_{mv}=f(I')-f(I)\\
\end{equation}
where $f$ is the optimization objective (see Formula (\ref{Fitness})).

In order to efficiently evaluate the move gain of a move, we develop dedicated techniques which are described in Section \ref{bucket_sort}. In this work, we employ two basic move operations: the \textit{'single-transfer move'} and the \textit{'double-transfer move'}. These two move operations form the basis of our five search operators.

\begin{itemize}
\item{Single-transfer move (st)}: Given a $k$-cut $I = \{S_1 , S_2, \ldots, S_k \}$, a vertex $v \in S_p$ and a target subset $S_q$ with $p,q \in \{1,\ldots,k\}, p\neq q$, the 'single-transfer move' displaces a single vertex $v \in S_p$ from its current subset $S_p$ to the target subset $S_q \neq S_p$. We denote this move by $st(v,S_p,S_q)$ or $v\rightarrow S_q$.

\item{Double-transfer move (dt)}: Given a $k$-cut $I = \{S_1 , S_2, \ldots, S_k \}$, the 'double-transfer move' displaces vertex $u$ from its subset $S_{cu}$ to a target subset $S_{tu} \neq  S_{cu}$, and displaces vertex $v$ from its current subset $S_{cv}$ to a target subset $S_{tv} \neq  S_{cv}$. We denote this move by $dt(u,S_{cu},S_{tu};v,S_{cv},S_{tv})$ or $dt(u,v)$, or still $dt$.
\end{itemize}

From these two basic move operations, we define five distinct \textit{search operators} $O_1 - O_5$ which indicate precisely how these two basic move operations are applied to transform an incumbent solution to a new solution. After an application of any of these search operators, the move gains of the impacted moves are updated according to the dedicated techniques explained in Section \ref{bucket_sort}.

\begin{itemize}
     \item  \textbf{The $\mathbf{O_1}$ search operator} applies the single-transfer move operation. Precisely, $O_1$ selects among the $(k-1)n$ single-transfer moves a best move $v\rightarrow S_q$ such that the induced move gain $\Delta_{(v\rightarrow S_q)}$ is maximum. If there are more than one such moves, one of them is selected at random. Since there are $(k-1)n$ candidate single-transfer moves from a given solution, the time complexity of $O_1$ is bounded by $O(kn)$. The proposed MOH algorithm employs this search operator as its main intensification operator which is  complemented by the $O_2$  search operator to locate good local optima.\\

    \item  \textbf{The $\mathbf{O_2}$ search operator} is based on the double transfer move operation and selects a best $dt$ move with the largest move gain $\Delta_{dt}$. If there are more than one such moves, one of them is selected at random.

        Let $dt(u,S_{cu},S_{tu};v,S_{cv},S_{tv})$ ($S_{cu}\neq S_{tu}$, $S_{cv}\neq S_{tv}$) be a double-transfer move, then the move gain $\Delta_{dt}$ of this double transfer move can be calculated by a combination of the move gains of its two underlying single-transfer moves ($\Delta_{u\rightarrow S_{tu}}$ and $\Delta_{v\rightarrow S_{tv}}$) as follows:
        \begin{equation}\label{eq-2mv-gain}
            \Delta_{dt(u,v)} = \Delta_{u\rightarrow S_{tu}} + \Delta_{v\rightarrow S_{tv}} + \psi\omega_{uv} \\
        \end{equation}
        where $\omega_{uv}$ is the weight of edge $e(u,v) \in E$ and $\psi$ is a coefficient which is determined as follows:
        \begin{equation}\label{eq-coe}
        \psi =
        \begin{cases}
      -2,  &\text{ if  } S_{cu} = S_{cv}, S_{tu} = S_{tv} \\
      -1, &\text{ if } S_{cu} = S_{cv}, S_{tu} \neq S_{tv}\\
      -1, &\text{ if } S_{cu} \neq S_{cv}, S_{tu} = S_{tv}\\
       1, &\text{ if  } S_{cu} \neq S_{cv}, S_{tu} = S_{cv}, S_{tv} \neq S_{cu}\\
       1, &\text{ if  } S_{cu} \neq S_{cv}, S_{tu} \neq S_{cv}, S_{tv} = S_{cu}\\
       2,  &\text{ if  } S_{cu} \neq S_{cv}, S_{tu} = S_{cv}, S_{tv} = S_{cu}\\
     0, &\text{ if } S_{cu} \neq S_{cv},  S_{tu} \neq S_{cv}, S_{tv} \neq S_{cu}, S_{tu} \neq S_{tv}
    \end{cases}
    \end{equation}
       It is clear that for a given incumbent solution, there are $(k-1)^2n^2$ candidate double-transfer moves denoted as set $DT$. Seeking directly the move with the maximum $\Delta_{dt}$ among all these possible moves would just be too computationally expensive. In order mitigate this problem, we devise a strategy to accelerate the move evaluation process.

        From Formula (\ref{eq-2mv-gain}), one observes that among all the vertices in $V$, only the vertices verifying the condition $\omega_{uv} \neq 0$ and $\Delta_{dt(u,v)} > 0$ are of interest for the double-transfer moves. Thus, by examining all the endpoint vertices of edges in $E$, we shrink the move combinations by building a reduced subset: $DT^R = \{st(v,S_p,S_q) \in \ DT: \exists st(u,S_{p'},S_{q'})  \in \ DT, \omega_{uv} \neq 0, \Delta_{dt(u,v)} > 0\}$. Based on $DT^R$, the complexity of examining all possible double-transfer moves drops to $O(|E|)$, which is not related to $k$. In practice, one can examine $\phi |E|$ endpoint vertices in case $|E|$ is too large. We empirically set $\phi = 0.1/d$, where $d$ is the highest degree of the graph.

        To summarize, the $O_2$ search operator selects two $st$ moves $u\rightarrow S_{tu}$ and $v\rightarrow S_{tv}$ from the reduced set $DT^R$, such that the combined move gain $\Delta_{dt(u,v)}$ according to Formula (\ref{eq-2mv-gain}) is maximum.

        Operator $O_2$ is used when $O_1$ exhausts its improving moves and provides a first means to help the descent-based improvement phase to escape the current local optimum and discover solutions of increasing quality.\\

    \item  Like $O_1$, \textbf{the $\mathbf{O_3}$ search operator} selects a best single-transfer move (i.e., with the largest move gain)  while considering a tabu list $H$ \cite{glover1999tabu}. The tabu list is a memory which is used to keep track of the performed $st$ moves to avoid revisiting previously encountered solutions. As such, each time a best $st$ move is performed to move a vertex $v$ from its original subset to a target subset, $v$ becomes tabu and is forbidden to move back to its original subset for the next $\lambda$ iterations (called tabu tenure), which is dynamically determined as follows.
        \begin{equation}\label{formula-tabu}
         \lambda = rand(3,n/10 )\\
        \end{equation}
        where $rand(3,n/10)$ denotes a random integer between $3$ and $n/10$. 

        Based on the tabu list, $O_3$ considers all possible single-transfer moves except those forbidden by the tabu list $H$ and selects the best $st$ move with the largest move main $\Delta_{st}$. Note that a forbidden move is always selected if the move leads to a solution better the best solution found so far. This is called aspiration in tabu search terminology \cite{glover1999tabu}.

        Operator $O_3$ is jointly used with operator $O_4$ to ensure the diversified improvement search.\\

    \item  Like $O_2$, \textbf{the $\mathbf{O_4}$ search operator} is based on the double-transfer operation. However, $O_4$ strongly constraints the considered candidate $dt$ moves with respect to two target subsets which are randomly selected. Specifically, $O_4$ operates as follows. Select two target subsets $S_p$ and $S_q$ at random, and then select two single-transfer moves $u \rightarrow S_p$ and $v\rightarrow S_q$ such that the combined move gain $\Delta_{dt(u,v)}$ according to Formula (\ref{eq-2mv-gain}) is maximum.

        Operator $O_4$ is jointly used with operator $O_3$ to ensure the diversified improvement search phase.\\

    \item   \textbf{The $\mathbf{O_5}$ search operator} is based on a randomized single-transfer move operation. $O_5$ first selects a random vertex $v \in V$ and a random target subset $S_p$, where $v \not\in S_p$ and then moves $v$ from its current subset to $S_p$. This operator is used to change randomly the incumbent solution for the purpose of (strong) diversification when the search is considered to be trapped in a deep local optimum.

\end{itemize}

Among the five search operators, four of them need to find a single-transfer move with the maximum move gain. To ensure a high computational efficiency of these operators, we develop below a streamlining technique for fast move gain evaluation and move gain updates.

\subsection{Bucket sorting for fast move gain evaluation and updating}\label{bucket_sort}
As mentioned in Section \ref{subsec_mov_ops}, to choose an appropriate move, it is crucial for the algorithm to be able to rapidly evaluate a number of candidate moves at each iteration. Since all the search operators basically rely on the single-transfer move operation, we devise a fast incremental evaluation technique based on a bucket data structure and a streamlining calculation to keep and update the move gains after each move application.

Our streamlining technique can be described as follows: let $v\rightarrow S_x$ be the move of transferring vertex $v$ from its current subset $S_{cu}$ to any other subset $S_x$, $x \in \{1,\dots,k\}, x \neq cu$. Then initially, each move gain can be determined by the following Formula:
\begin{equation}\label{form-init-move-gain}
\small
    \Delta_{v\rightarrow S_x} =\sum\limits_{i\in S_{cu}, i\neq v}\omega_{vi} - \sum\limits_{j\in S_x} \omega_{vj}, \ x \in \{1,\dots,k\}, x \neq cu\\
\end{equation}
where $\omega_{vi}$ and $\omega_{vj}$ are respectively the weights of edges $e(v,i)$ and $e(v,j)$.

Suppose the move $v\rightarrow S_{tv}$, i.e., displacing $v$ from its current subset $S_{cv}$ to target subset $S_{tv}$,  is performed, the algorithm needs to update the move gains by performing the following calculation:

\begin{enumerate}

  \item  $\Delta_{v\rightarrow S_{cv}} = - \Delta_{v\rightarrow S_{dv}}$
  \item  $\Delta_{v\rightarrow S_{dv}} = 0$
  \item  for each $S_x\neq S_{cv}$, $S_x\neq S_{dv}$, \\
         $\Delta_{v\rightarrow S_x} = \Delta_{v\rightarrow S_x}$
  \item  for each $u\in V-\{v\}$,  moving $u\in S_{cu}$ to each other subset $S_y \in S-\{S_{cu}\}$,\\
  \begin{equation}\label{eq_upd_gain}
    \Delta_{u\rightarrow S_y} =
    \begin{cases}
     \Delta_{u\rightarrow S_y} - 2\omega_{uv}, &\text{ if } S_{cu} = S_{cv}, S_y = S_{dv} \\
     \Delta_{u\rightarrow S_y}, &\text{ if } S_{cu} = S_{cv}, S_y \neq S_{dv}, S_y\neq S_{cv} \\
     \Delta_{u\rightarrow S_y} + 2 \omega_{uv}, &\text{ if } S_{cu} = S_{dv}, S_y = S_{cv} \\
     \Delta_{u\rightarrow S_y}, &\text{ if } S_{cu} = S_{dv}, S_y\neq S_{cv} \\
     \Delta_{u\rightarrow S_y} +  \omega_{uv}, &\text{ if } S_{cu} \neq S_{cv}, S_{cu} \neq S_{dv}, S_y = S_{cv} \\
     \Delta_{u\rightarrow S_y} -  \omega_{uv}, &\text{ if } S_{cu} \neq S_{cv}, S_{cu} \neq S_{dv}, S_y = S_{dv} \\

    \end{cases}
  \end{equation}
\end{enumerate}

It is easy to see that only the move gains of vertices affected by this move (i.e., the displaced vertex and its adjacent vertices) will be updated, which reduces the computation time significantly.

Normally, the move gains can be stored in an array, with which the time for finding the best move with maximum move gain grows linearly with the number of vertices ($O(n))$. For large problem instances (very large $n$), the required time can still be quite high. To avoid unnecessary searching for the vertex for the best move, we adopts a bucket structure which keeps vertices ordered by their move gains in decreasing order. The bucket sorting was proposed by Fiduccia and Mattheyes to improve the Kerninghan-Lin heuristic for the network partitioning problem \cite{fiduccia1982linear}.  In this work, we adapt for the first time the idea of bucket sorting for the max-k-cut problem.  This is done by using $k$ arrays of buckets, one for each partition subset $S_{i}\in \{S_{1},S_{2},\ldots,S_{k}\}$. In each bucket array $i$, $1\leq i\leq k$, the $j^{th}$ entry stores the vertices with the move gain $\Delta_{v\rightarrow S_i}$ currently equaling $j$, where those vertices are maintained by a doubly linked list. To ensure a direct access to the vertex in the doubly linked lists, as suggested in \cite{fiduccia1982linear}, the algorithm also maintains another array for all vertices, where each element points to its corresponding vertex in the doubly linked lists.

\begin{figure*}[h,t]
\centering
\hspace*{\fill}{\centering\includegraphics[width=5.3in]{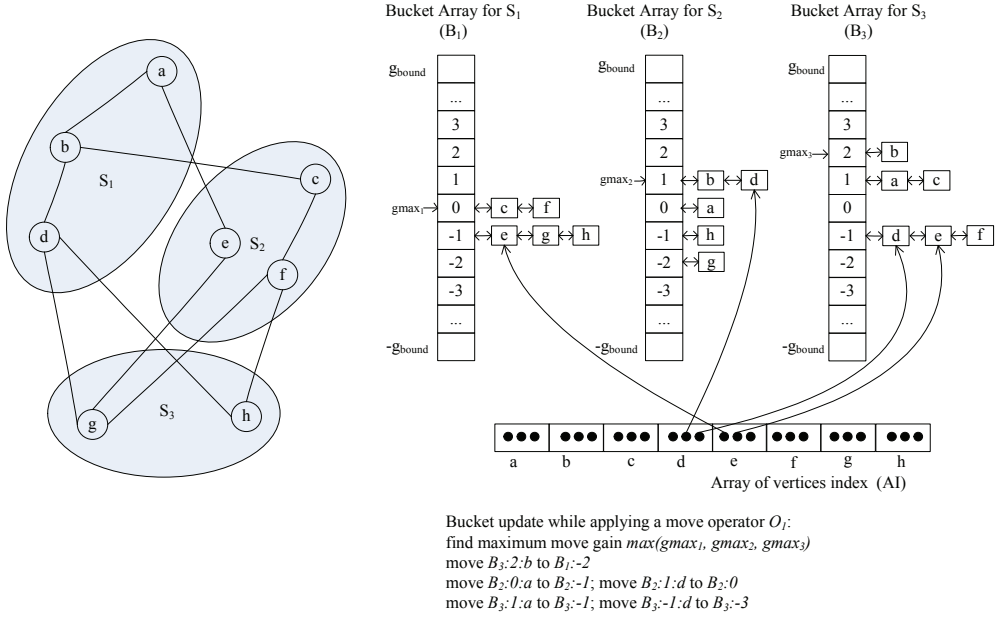}}
\centering\caption{An example of bucket structure for max-3-cut}\label{fig_bucket}
\end{figure*}

Fig. \ref{fig_bucket} shows an illustrative example of the bucket structure for max-k-cut for $k=3$. In the graph of the example (Fig. \ref{fig_bucket}, left), there are a total of 8 vertices belonging to the 3 subsets $S_1, S_2$ and $S_3$. The bucket structure for this graph is shown Fig. \ref{fig_bucket} (right). From the graph, one observes that the move gains of moving vertices $e,g,h$ to subset $S_1$ equal $-1$, then those three vertices are stored in the entry of $B_1$ with index of -1. Notice that vertices $e,g,h$ are managed as a doubly linked list. The array AI shown at the bottom of  Fig. \ref{fig_bucket} manages position indexes of for all vertices.

For each array of buckets, finding the best vertex with maximum move gain is equivalent to finding the first non-empty bucket from top of the array and then selecting a vertex in its doubly linked list. If there are more than one vertices in the doubly linked list, a random vertex in this list is selected. To further reduce the searching time, the algorithm memorizes the position of the first non-empty bucket (e.g., $gmax_1, gmax_2, gmax_3$ in Fig. \ref{fig_bucket}).

After each move, the bucket structure is updated by recomputing the move gains (see Formula (\ref{eq_upd_gain})) of the affected vertices which include the moved vertex and its adjacent vertices, and shifting them to appropriate buckets.

For instance, the steps of performing an $O_1$ move based on Fig. \ref{fig_bucket} are shown as follows: First, obtain the index of maximum move gain in the bucket arrays by calculating $max (gmax_1, gmax_2, gmax_3)$, which equals $gmax_3$ in this case. Second, select randomly a vertex indexed by $gmax_3$, vertex $b$ in this case. At last, update the positions of the affected vertices $a, b, d$.

The complexity of each move consists in searching for the vertex with maximum move gain, recomputing the move gain for the affected vertices and updating the bucket structure. The vertex with maximum move gain can be simply obtained in constant time.  Recomputing move gain is in linear time relative to the number of affected vertices. The time of updating the bucket structure is also only related to the number of affected vertices. As a result, $k$ has no influence on the performance of the proposed algorithm in terms of computing time. However, it does require a greater amount of memory as $k$ increases.

\subsection{Descent-based improvement phase for intensified search}\label{subsec_descent_search}
The descent-based local search is used to obtain a local optimum from a given starting solution. As described in Algorithm \ref{algo_mkc} (lines $11$ - $21$), we alternatively uses two search operators $O_1$ and $O_2$ defined in Section \ref{subsec_mov_ops} to improve a solution until reaching a local optimum. Starting from the given initial solution, the procedure progressively applies $O_1$ to the incumbent solution. According to the definition of $O_1$ in Section \ref{subsec_mov_ops}, at each step, the procedure examines all possible single-transfer moves and selects a move $v\rightarrow S_q$ with the largest move gain $\Delta_{v\rightarrow S_q}$ subject to $\Delta_{v\rightarrow S_q}>0$, and then performs that move. After the move, the algorithm updates the bucket structure of move gains according to the technique described in Section \ref{bucket_sort}. 

When the incumbent solution can not be improved by the $O_1$ operator (i.e., $\forall v\in V, \forall S_q, \Delta_{v\rightarrow S_q} \leq 0$), the procedure turns to operator $O_2$ which makes one \textit{best} double-transfer move. If an improved solution is discovered with respect to the local optimum reached by $O_1$, we are in a new promising area. We switch back to operator $O_1$ to resume an intensified search. The descent-based improvement phase stops when no better solution can be found with $O_1$ and $O_2$. This solution is a local optimum $I_{lo}$ with respect to the single-transfer and double-transfer moves and serves as the input solution of the second search phase which is explained in the next section.

\subsection{Diversified improvement phase for discovering promising region}\label{subsec_div_search}
The descent-based local phase described in Section \ref{subsec_descent_search} alone can not escape the best local optimum $I_{lo}$ it encounters. The diversified improvement search phase is used 1) to jump out of this local optimum and 2) to intensify the search around this local optimum with the hope of discovering a solution better than the input local optimum $I_{lo}$.

The diversified improvement search procedure alternatively uses two search operators $O_3$ and $O_4$ defined in Section \ref{subsec_mov_ops} to perform a move until the stop criterion is met. The application of $O_3$ or $O_4$ is determined  probabilistically: with probability $\rho$, $O_3$ is applied; with $1-\rho$, $O_4$ is applied.

When $O_3$ is selected, the algorithm searches for a best single transfer move $v\rightarrow S_q$ with maximum move gain $\Delta_{v\rightarrow S_q}>0$ which is not forbidden by the tabu list or verifies the aspiration criterion. Each performed move is then recorded in the tabu list $H$ and is classified tabu for the next $\lambda$ (calculated by Formula (\ref{formula-tabu})) iterations. The bucket structure is updated to actualize the impacted move gains accordingly. Note that the algorithm only keeps and updates the tabu list during the diversified improvement search phase. Once this second search phase terminates, the tabu list is cleared up.

Similarly, when $O_4$ is selected, two subsets are selected at random and a best double-transfer $dt$ move with maximum move gain $\Delta_{dt}$ is determined from the bucket structure. After the move, the bucket structure is updated to actualize the impacted move gains. It is notated that in case of multiple best double-transfer moves, one of them is chosen at random.

The diversified improvement search procedure terminates once a  solution better than the input local optimum $I_{lo}$ is found, or a maximum number $\omega$ of diversified moves ($O_3$ or $O_4$) is reached. Then the algorithm returns to the descent-based search procedure and use the current solution $I$ as a new starting point for the descent-based search. If the best solution founded so far ($f_{best}$) can not be improved over a maximum allowed number $\xi$ of consecutive rounds of the descent-based improvement and diversified improvement phases, the perturbation phase (Section \ref{subsec_div_pertub}) is invoked  to displace the search to a distant region.

\subsection{Perturbation phase for strong diversification}\label{subsec_div_pertub}
The diversified improvement phase makes it possible for the search to escape some local optima. However, the algorithm may still get deeply stuck in a non-promising regional search area. This is the case when the best-found solution $f_{best}$ can not be improved after $\xi$ consecutive rounds of descent and diversified improvement phases. Thus the random perturbation is developed to displace the search into a more distant region.

The basic idea of the perturbation consists in applying the $O_5$ operator $\gamma$ times. In other words, this perturbation phase moves $\gamma$ randomly selected vertices from their original subset to a new and randomly selected subset. Here, $\gamma$ is used to control the perturbation strength; a large (resp. small)  $\gamma$ value changes strongly (resp. weakly) the incumbent solution. In our case, we adopt $\gamma=0.1|V|$, i.e., as a percent of the number of vertices. After the perturbation phase, the search returns to the descent-based improvement phase with the perturbed solution as its new starting solution.

\section{Experimental results and comparisons}\label{sec_Results}
\subsection{Benchmark instances}\label{subsec_Benchmarks}
To evaluate the performance of the proposed MOH approach, we carry out computational experiments on two sets of well-known benchmarks with a total of 91 large instances of the literature\footnote{Our best results are  available at: \url{http://www.info.univ-angers.fr/pub/hao/maxkcut/MOHResults.zip}.}. The first set (G-set) is composed of 71 graphs with 800 to 20000 vertices and an edge density from 0.02\% to 6\%. These instances are generated by a machine-independent graph generator including toroidal, planar and random weighted graphs. These instances are available from: http://www.stanford.edu/yyye/yyye/Gset. The second set comes form  \cite{burer2002rank}, arising from 30 cubic lattices with randomly generated interaction magnitudes. Since the 10 small instances (with less than 1000 vertices) are very easy for our algorithm, only the results of the 20 larger instances with 1000 to 2744 vertices are reported. These well-known benchmarks are frequently used to evaluate the performance of max-bisection, max-cut and max-k-cut algorithms \cite{benlic2013breakout,festa2002randomized,shylo2012solving,shyloGlover2015,wang2013probabilistic,wu2012memetic,wu2013memetic,Zhu2013}.

\subsection{Experimental protocol}\label{subsec_Protocol}
Our MOH algorithm is programmed in C++ and compiled with GNU g++ (optimization flag "-O2"). Our computer is equipped with a Xeon E5440/2.83GHz CPU with 2GB RAM. When testing the DIMACS machine benchmark\footnote{dfmax:\url{ftp://dimacs.rutgers.edu/pub/dsj/clique/}}, our machine requires 0.43, 2.62 and 9.85 CPU time in seconds respectively for graphs r300.5, r400.5, and r500.5 compiled with g++ -O2.

\subsection{Parameters}\label{subsec_para}
The proposed algorithm requires several parameters: tabu tenure $\lambda$, max allowed number $\omega$ of consecutive TS moves, max allowed number $\xi$ of consecutive rounds of descent improvement and diversified improvement phases, probability $\rho$ for selecting tabu-based move operator $O_3$, and number $\gamma$ of perturbation moves. The parameter values were determined by performing a preliminary experiment on a selection of 10 representative and challenging instances from the G-set benchmark: G22, G23, G25, G29, G33, G35, G36, G37, G38, G40. For each parameter, we tested a range of different values, while keeping the rest of the parameters to their default values. To report our computational results, we adopt the set of parameter values $(\lambda=rand(3,|V|/10 ), \omega=500, \xi=1000 , \rho=0.5, \gamma=0.1|V|)$ for all our experiments throughout the paper, though it would be possible to attain better results by further fine-turning the parameters.

Considering the stochastic nature of our MOH algorithm, each instance is independently solved multiple times: 20 times for max-cut ($k=2$), 10 times for max-k-cut ($k>2$). For the purpose of fair comparisons reported in Sections \ref{subsec-comp-k-cut} and \ref{subsec-comp-2-cut}, we follow the reference algorithms and use a timeout limit as the stop criterion of our MOH algorithm. The timeout limit is set to be 30 minutes for graphs with $|V| < 5000$, 120 minutes for graphs with $10000\geq |V| \geq 5000$, 240 minutes for graphs with $|V| \geq 10000$.

To fully evaluate the performance of the proposed algorithm, we investigate two comparisons with the state of the art algorithms. First, we focus on the max-k-cut problem ($k=2,3,4,5$), where we thoroughly compare our algorithm with the recent discrete dynamic convexized algorithm  \cite{Zhu2013} which provides the most competitive results for the general max-k-cut problem in the literature. Secondly, for the special max-cut case ($k=2$), we compare our algorithm with six most recent max-cut algorithms \cite{benlic2013breakout,kochenberger2013solving,shylo2012solving,wang2013probabilistic,wu2012memetic,wu2013memetic}. It should be noted that those state of the art max-cut algorithms are specifically designed the particular max-cut problem while our algorithm is developed for the general max-k-cut problem. Normally, the dedicated algorithms are advantaged since they can better explore the particular features of the max-cut problem.

\subsection{Comparison with state-of-the-art max-k-cut algorithms}\label{subsec-comp-k-cut}
In this section, we present the results attained by our MOH algorithm for the max-k-cut problem. As mentioned above, we compare the proposed algorithm with the discrete dynamic convexized algorithm (DC) \cite{Zhu2013}, which was published very recently. DC was tested on a computer with a 2.11 GHz AMD processor and 1 GB of RAM. According to the Standard Performance Evaluation Cooperation (SPEC) (www.spec.org), this computer is 1.4 times slower than the computer we used for our experiments. Note that DC is the only heuristic algorithm available in the literature, which published computational results for the general max-k-cut problem.

Tables \ref{table-comp-2cut} to \ref{table-comp-5cut} respectively show the computational results of our MOH algorithm ($k=2,3,4,5$) on the 2 sets of benchmarks in comparison with those of the DC algorithm. The first two columns of the tables indicate the name and the number of vertices of the graphs. Columns 3 to 7 present the results attained by our algorithm, where $f_{best}$ and $f_{avg}$ show the best objective value and the average objective value over 20 runs, $std$ gives the standard deviation and $time(s)$ indicates the average CPU time in seconds required by our algorithm to reach the best objective value $f_{best}$. Columns 8 to 9 present the results ($f_{best}$, $f_{avg}$, $time(s)$) attained by DC. Considering the difference between our computer and the computer used by DC, we normalize the $times$ of DC by dividing them by 1.4 according to the SPEC mentioned above. The entries marked as "-" in the tables indicate that the corresponding results are not available. The entries in bold indicate that those results are better than the results provided by the reference DC algorithm. The last column gives the gaps $gap$ of the best objective value for each instance between our algorithm and DC. A positive gap implies an improved result.

From Table \ref{table-comp-2cut} on max-2-cut, one observes that our algorithm achieves a better $f_{best}$ (best objective value) for 50 out of 74 instances reported by DC, while a better $f_{avg}$ (average objective value) for 71 out of 74 instances. Our algorithm matches the results on other instances and there is no result worse than that obtained by DC. The average standard deviation for all 91 instances is only 2.82, which shows our algorithm is stable and robust.

From Table \ref{table-comp-3cut}, \ref{table-comp-4cut}, and \ref{table-comp-5cut},  which respectively show the comparative results on max-3-cut, max-4-cut and max-5-cut. One observes that our algorithm achieves much higher solution quality on more than 90 percent of 44 instances reported by DC while getting 0 worse result. Moreover, even our \textit{average} results ($f_{avg}$) are  better than the \textit{best} results reported by DC.

Our algorithm is also highly competitive in terms of computing time. It is not fully fair to directly compare the columns $times$  for the two algorithms, because the $times$ indicate the average time needed for the algorithm to attain its $f_{best}$ value, while the $f_{best}$ values obtained by the two algorithms are different. One observes that for most cases, our algorithm consumes significantly less time while obtaining better results, indicating that our algorithm can reach better solutions with less computing times. This is particularly true when $k>2$.

We conclude that the proposed algorithm for the general max-k-cut probleme dominates the state of the art reference DC algorithm both in terms of solution quality and computing time.

\subsection{Comparison with state-of-the-art max-cut algorithms}\label{subsec-comp-2-cut}
Our algorithm is designed for the general max-k-cut problem for $k\geq2$. The assessment of the last section focuses on the general case. In this section, we further evaluate the performance of the proposed algorithm for the special max-cut problem ($k=2$).

Recall that max-cut has been largely studied in the literature for a long time and there are many powerful heuristics which are specifically designed for the problem. These state-of-the-art max-cut algorithms constitute thus relevant references for our comparative study. In particular, we adopt the following 6 best performing sequential algorithms published since 2012.

\begin{enumerate}
  \item Global equilibrium search (GES) \cite{shylo2012solving} - an algorithm sharing ideas similar to simulated annealing and utilizing accumulated information of search space to generate new solutions for the subsequent stages. The reported results of GES were obtained on a PC with a 2.83GHz Intel Core QUAD Q9550 CPU and 8.0GB RAM.
  \item Breakout local search (BLS) \cite{benlic2013breakout} - a heuristic algorithm integrating a local search and adaptive perturbation strategies. The reported results of BLS were obtained on a PC with 2.83GHz Intel Xeon E5440 CPU and 2GB RAM.
  \item Two memetic algorithms respective for the max-cut problem (MACUT) \cite{wu2012memetic} and the max-bisection problem (MAMBP) \cite{wu2013memetic} - integrating a grouping crossover operator and a tabu search procedure. The results reported in the two papers were obtained on a PC with a 2.83GHz Intel Xeon E5440 CPU and 2GB RAM.
  \item GRASP-Tabu search algorithm \cite{wang2013probabilistic} - a method converting the max-cut problem to the UBQP problem and solving it by integrating GRASP and tabu search. The reported results were obtained on a PC with a 2.83GHz Intel Xeon E5440 CPU and 2GB RAM.
  \item Tabu search (TS-UBQP) \cite{kochenberger2013solving} - a tabu search algorithm designed for UBQP. The evaluation of TS-UBQP  were performed on a PC with a 2.83GHz Intel Xeon E5440 CPU and 2GB RAM.

\end{enumerate}

One notices that except GES, the other five reference algorithms were run on the same computer platform. Nevertheless, it is still difficult to make a fully fair comparison of the computing time, due to the differences on programming language, compiling options, and termination conditions, etc. Our comparison thus focuses on the best solution achieved by each algorithm. Recall that for our algorithm, the timeout limit is set to be 30 minutes for graphs with $|V| < 5000$, 120 minutes for graphs with $1000\geq |V| \geq 5000$, 240 minutes for graphs with $|V| \geq 10000$. Our algorithm employs thus the same timeout limits as \cite{wu2012memetic} on the graphs $|V|< 10000 $, but for the graphs $ |V| \geq 10000$, we use 240 minutes to compare with BLS \cite{benlic2013breakout}. 

Table \ref{table-comp-5ref} gives the comparative results on the 91 instances of the two benchmarks. Columns 1 and 2 respectively indicate the instance name and the number of vertices of the graphs.  Columns 3 shows the current best known objective value $f_{pre}$ reported by any existing max-cut algorithm in the literature including the latest \textit{parallel} GES algorithm \cite{shyloGlover2015}. Columns 4 to 9 give the best objective value obtained by the 6 reference algorithms: GES \cite{shylo2012solving}, BLS \cite{benlic2013breakout}, MACUT \cite{wu2012memetic}, TS-UBQP \cite{kochenberger2013solving}, GRASP-TS/PM \cite{wang2013probabilistic}, MAMBP \cite{wu2013memetic}. Note that MAMBP is designed for the max-bisection problem (i.e., balanced max-cut), however it achieves some previous best known max-cut results.  The last column 'MOH' recalls the best results of our algorithm from Table \ref{table-comp-2cut}. The rows denoted by 'Better', 'Equal' and 'Worse' respectively indicate the number of instances for which our algorithm obtains a result of better, equal and worse quality relative to each reference algorithm. The entries are reported in the form of x/y/z, where z denotes the total number of the instances tested by our algorithm, y is the number of the instances tested by a reference algorithm and x indicates the number of instances where our algorithm achieved 'Better', 'Equal' or 'Worse' results. The results in bold mean that our algorithm has improved the best known results. The entries marked as "-" in the table indicate that the results are not available.

From Table \ref{table-comp-5ref}, one observes that our algorithm is able to improve the current best known results in the literature for 6 instances, and match the best known results for 73 instances.  For 12 cases (in italic), even if our results are worse than the current best known results achieved by the latest \textit{parallel} GES algorithm \cite{shyloGlover2015}, they are still better than the results of any other existing algorithms including TS-UBQP \cite{wang2013probabilistic} and BLS \cite{benlic2013breakout}. Note that the results of the parallel GES algorithm are achieved on a more powerful computing platform (Intel CoreTM i7-3770 CPU @3.40 GHz and 8.0 GB RAM) and with longer time limits (4 parallel processes at the same time and 1 hour for each process). 

Such a performance is remarkable given that we are comparing our more general algorithm designed for max-k-cut with the best performing specific max-cut algorithms. The experimental evaluations presented in this section and last section demonstrate that our algorithm not only performs well on the general max-k-cut problem, but also remains highly competitive for the special case of the popular max-cut problem.

\newpage
{\renewcommand{\arraystretch}{0.8}
{\tiny
\begin{longtable}{@{}lrrrrrrrrrrrr@{}}
\caption{Comparative results for max-2-cut between the proposed MOH algorithm and DC \cite{Zhu2013}. }\label{table-comp-2cut}\\
\toprule
    Instance &     $|V|$ &    \multicolumn{ 4}{c}{MOH} &   & \multicolumn{ 3}{c}{DC} &  & $gap$\\ \cmidrule{3-6} \cmidrule{8-10}
   &   &   $f_{best}$&   $f_{avg}$&   $std$&   $time(s)$ &   & $f_{best}$ &  $f_{avg}$&  $time(s)$ &  &\\
\midrule
\endfirsthead

\multicolumn{13}{c}%
{{\bfseries \tablename\ \thetable{} -- continued from previous page}} \\
\toprule
    Instance &     $|V|$ &    \multicolumn{ 4}{c}{MOH} &   & \multicolumn{ 3}{c}{DC} &  & $gap$  \\\cmidrule{3-6} \cmidrule{8-10} &  & $f_{best}$&   $f_{avg}$&   $std$&   $time(s)$ &   & $f_{best}$ &  $f_{avg}$&  $time(s)$ &  &  \\
\midrule
\endhead
\midrule
\endfoot
\hline \hline
\endlastfoot

G1    & 800   & 11624 & \textbf{11624.00 } & 0.00  & 1.5   &       & 11624 & 11617.20  & 131.7  &       & 0 \\
G2    & 800   & 11620 & \textbf{11620.00 } & 0.00  & 4.6   &       & 11620 & 11610.00  & 131.4  &       & 0 \\
G3    & 800   & 11622 & \textbf{11622.00 } & 0.00  & 1.2   &       & 11622 & 11612.20  & 130.8  &       & 0 \\
G4    & 800   & 11646 & \textbf{11646.00 } & 0.00  & 5.2   &       & 11646 & 11633.90  & 133.8  &       & 0 \\
G5    & 800   & 11631 & \textbf{11631.00 } & 0.00  & 1.0   &       & 11631 & 11623.20  & 131.7  &       & 0 \\
G6    & 800   & 2178  & \textbf{2178.00 } & 0.00  & 3.0   &       & 2178  & 2175.90  & 132.1  &       & 0 \\
G7    & 800   & 2006  & \textbf{2006.00 } & 0.00  & 3.0   &       & 2006  & 1997.70  & 137.6  &       & 0 \\
G8    & 800   & 2005  & \textbf{2005.00 } & 0.00  & 5.7   &       & 2005  & 2000.00  & 139.2  &       & 0 \\
G9    & 800   & \textbf{2054} & \textbf{2054.00 } & 0.00  & 3.2   &       & 2049  & 2043.50  & 134.9  &       & 5 \\
G10   & 800   & \textbf{2000} & \textbf{2000.00 } & 0.00  & 68.1  &       & 1999  & 1998.40  & 133.3  &       & 1 \\
G11   & 800   & 564   & \textbf{564.00 } & 0.00  & 0.2   &       & 564   & 563.80  & 58.8  &       & 0 \\
G12   & 800   & 556   & \textbf{556.00 } & 0.00  & 3.5   &       & 556   & 555.40  & 58.7  &       & 0 \\
G13   & 800   & 582   & \textbf{582.00 } & 0.00  & 0.9   &       & 582   & 580.00  & 60.9  &       & 0 \\
G14   & 800   & \textbf{3064} & \textbf{3064.00 } & 0.00  & 251.3  &       & 3057  & 3054.30  & 82.7  &       & 7 \\
G15   & 800   & \textbf{3050} & \textbf{3050.00 } & 0.00  & 52.2  &       & 3044  & 3038.00  & 82.4  &       & 6 \\
G16   & 800   & 3052  & \textbf{3052.00 } & 0.00  & 93.7  &       & 3052  & 3039.60  & 81.1  &       & 0 \\
G17   & 800   & \textbf{3047} & \textbf{3047.00 } & 0.00  & 129.5  &       & 3043  & 3037.80  & 81.6  &       & 4 \\
G18   & 800   & \textbf{992} & \textbf{992.00 } & 0.00  & 112.6  &       & 989   & 984.00  & 89.1  &       & 3 \\
G19   & 800   & 906   & \textbf{906.00 } & 0.00  & 266.9  &       & 906   & 899.90  & 84.4  &       & 0 \\
G20   & 800   & 941   & \textbf{941.00 } & 0.00  & 43.7  &       & 941   & 938.20  & 86.3  &       & 0 \\
G21   & 800   & 931   & \textbf{931.00 } & 0.00  & 155.3  &       & 931   & 926.00  & 86.2  &       & 0 \\
G22   & 2000  & \textbf{13359} & \textbf{13357.00 } & 1.91  & 352.4  &       & 13339 & 13315.90  & 683.7  &       & 20 \\
G23   & 2000  & \textbf{13344} & \textbf{13344.00 } & 0.00  & 433.8  &       & 13323 & 13298.90  & 705.2  &       & 21 \\
G24   & 2000  & \textbf{13337} & \textbf{13336.70 } & 0.46  & 777.9  &       & 13314 & 13286.00  & 692.1  &       & 23 \\
G25   & 2000  & \textbf{13340} & \textbf{13335.50 } & 2.40  & 442.5  &       & 13324 & 13293.70  & 694.7  &       & 16 \\
G26   & 2000  & \textbf{13328} & \textbf{13325.50 } & 2.31  & 535.1  &       & 13313 & 13282.20  & 689.6  &       & 15 \\
G27   & 2000  & \textbf{3341} & \textbf{3341.00 } & 0.00  & 42.2  &       & 3326  & 3285.40  & 677.9  &       & 15 \\
G28   & 2000  & \textbf{3298} & \textbf{3298.00 } & 0.00  & 707.2  &       & 3292  & 3272.00  & 680.5  &       & 6 \\
G29   & 2000  & \textbf{3405} & \textbf{3397.85 } & 5.31  & 555.2  &       & 3390  & 3357.20  & 693.4  &       & 15 \\
G30   & 2000  & \textbf{3413} & \textbf{3412.15 } & 0.36  & 1427.0  &       & 3398  & 3369.50  & 676.5  &       & 15 \\
G31   & 2000  & \textbf{3310} & \textbf{3307.85 } & 0.91  & 592.6  &       & 3295  & 3273.90  & 696.4  &       & 15 \\
G32   & 2000  & \textbf{1410} & \textbf{1410.00 } & 0.00  & 65.7  &       & 1408  & 1402.70  & 514.9  &       & 2 \\
G33   & 2000  & \textbf{1382} & \textbf{1381.60 } & 0.80  & 504.1  &       & 1378  & 1373.70  & 508.8  &       & 4 \\
G34   & 2000  & \textbf{1384} & \textbf{1384.00 } & 0.00  & 84.2  &       & 1378  & 1376.70  & 531.5  &       & 6 \\
G35   & 2000  & \textbf{7687} & \textbf{7681.65 } & 1.59  & 796.7  &       & 7647  & 7632.20  & 614.5  &       & 40 \\
G36   & 2000  & \textbf{7680} & \textbf{7673.60 } & 1.62  & 1553.2  &       & 7625  & 7618.50  & 613.1  &       & 55 \\
G37   & 2000  & \textbf{7691} & \textbf{7685.75 } & 2.26  & 1195.1  &       & 7640  & 7627.70  & 623.7  &       & 51 \\
G38   & 2000  & \textbf{7688} & \textbf{7683.60 } & 2.27  & 30.6  &       & 7641  & 7614.40  & 632.9  &       & 47 \\
G39   & 2000  & \textbf{2408} & \textbf{2405.35 } & 1.85  & 787.7  &       & 2375  & 2352.50  & 659.3  &       & 33 \\
G40   & 2000  & \textbf{2400} & \textbf{2397.35 } & 2.43  & 472.5  &       & 2384  & 2371.70  & 656.8  &       & 16 \\
G41   & 2000  & \textbf{2405} & \textbf{2405.00 } & 0.00  & 377.3  &       & 2377  & 2357.40  & 666.8  &       & 28 \\
G42   & 2000  & \textbf{2481} & \textbf{2476.35 } & 2.01  & 65.1  &       & 2469  & 2441.30  & 657.1  &       & 12 \\
G43   & 1000  & \textbf{6660} & \textbf{6660.00 } & 0.00  & 1.2   &       & 6657  & 6648.90  & 156.7  &       & 3 \\
G44   & 1000  & 6650  & \textbf{6650.00 } & 0.00  & 5.3   &       & 6650  & 6643.70  & 155.8  &       & 0 \\
G45   & 1000  & \textbf{6654} & \textbf{6654.00 } & 0.00  & 6.9   &       & 6647  & 6640.70  & 155.3  &       & 7 \\
G46   & 1000  & \textbf{6649} & \textbf{6648.90 } & 0.30  & 67.3  &       & 6647  & 6637.90  & 157.0  &       & 2 \\
G47   & 1000  & 6657  & \textbf{6657.00 } & 0.00  & 43.3  &       & 6657  & 6648.50  & 157.8  &       & 0 \\
G48   & 3000  & 6000  & 6000.00  & 0.00  & 0.0   &       & 6000  & 6000.00  & 420.1  &       & 0 \\
G49   & 3000  & 6000  & 6000.00  & 0.00  & 0.0   &       & 6000  & 6000.00  & 440.3  &       & 0 \\
G50   & 3000  & 5880  & 5880.00  & 0.00  & 532.1  &       & 5880  & 5880.00  & 552.5  &       & 0 \\
G51   & 1000  & \textbf{3848} & \textbf{3848.00 } & 0.00  & 189.2  &       & 3842  & 3831.50  & 137.6  &       & 6 \\
G52   & 1000  & \textbf{3851} & \textbf{3851.00 } & 0.00  & 209.7  &       & 3840  & 3830.50  & 132.7  &       & 11 \\
G53   & 1000  & \textbf{3850} & \textbf{3849.95 } & 0.22  & 299.3  &       & 3844  & 3835.00  & 136.3  &       & 6 \\
G54   & 1000  & \textbf{3852} & \textbf{3851.10 } & 0.30  & 190.4  &       & 3831  & 3824.40  & 136.0  &       & 21 \\
G55   & 5000  & 10299 & 10283.40  & 7.13  & 1230.4  &       & -     & -     & -     &       & - \\
G56   & 5000  & 4016  & 4007.47  & 6.49  & 990.4  &       & -     & -     & -     &       & - \\
G57   & 5000  & 3494  & 3486.80  & 2.45  & 1528.3  &       & -     & -     & -     &       & - \\
G58   & 5000  & 19288 & 19275.40  & 4.58  & 1522.3  &       & -     & -     & -     &       & - \\
G59   & 5000  & 6087  & 6077.19  & 7.90  & 2498.8  &       & -     & -     & -     &       & - \\
G60   & 7000  & 14190 & 14173.00  & 6.98  & 2945.4  &       & -     & -     & -     &       & - \\
G61   & 7000  & 5798  & 5782.67  & 5.72  & 6603.3  &       & -     & -     & -     &       & - \\
G62   & 7000  & 4868  & 4851.73  & 7.10  & 5568.6  &       & -     & -     & -     &       & - \\
G63   & 7000  & 27033 & 27019.20  & 6.72  & 6492.1  &       & -     & -     & -     &       & - \\
G64   & 7000  & 8747  & 8700.87  & 17.28  & 4011.1  &       & -     & -     & -     &       & - \\
G65   & 8000  & 5560  & 5554.40  & 2.73  & 4709.5  &       & -     & -     & -     &       & - \\
G66   & 9000  & 6360  & 6354.53  & 2.37  & 6061.9  &       & -     & -     & -     &       & - \\
G67   & 10000 & 6942  & 6936.53  & 2.88  & 14214.3  &       & -     & -     & -     &       & - \\
G70   & 10000 & 9544  & 9527.80  & 9.93  & 6364.0  &       & -     & -     & -     &       & - \\
G72   & 10000 & 6998  & 6991.53  & 2.67  & 6586.6  &       & -     & -     & -     &       & - \\
G77   & 14000 & 9928  & 9920.00  & 3.08  & 9863.6  &       & -     & -     & -     &       & - \\
G81   & 20000 & 14036 & 14020.30  & 8.50  & 10422.0  &       & -     & -     & -     &       & - \\
3dl101000 & 1000  & 896   & \textbf{896.00 } & 0.00  & 4.4   &       & 896   & 888.70  & 113.3  &       & 0 \\
3dl102000 & 1000  & 900   & \textbf{900.00 } & 0.00  & 6.8   &       & 900   & 898.50  & 111.5  &       & 0 \\
3dl103000 & 1000  & \textbf{892} & \textbf{892.00 } & 0.00  & 147.5  &       & 888   & 884.70  & 113.0  &       & 4 \\
3dl104000 & 1000  & 898   & \textbf{898.00 } & 0.00  & 2.7   &       & 898   & 895.00  & 112.2  &       & 0 \\
3dl105000 & 1000  & \textbf{886} & \textbf{886.00 } & 0.00  & 11.7  &       & 884   & 882.80  & 115.0  &       & 2 \\
3dl106000 & 1000  & 888   & \textbf{888.00 } & 0.00  & 2.1   &       & 888   & 883.70  & 114.7  &       & 0 \\
3dl107000 & 1000  & \textbf{900} & \textbf{899.60 } & 1.00  & 42.9  &       & 898   & 892.40  & 114.1  &       & 2 \\
3dl108000 & 1000  & \textbf{882} & \textbf{882.00 } & 0.00  & 8.0   &       & 880   & 877.70  & 120.0  &       & 2 \\
3dl109000 & 1000  & \textbf{902} & \textbf{902.00 } & 0.00  & 18.7  &       & 902   & 894.40  & 113.6  &       & 0 \\
3dl1010000 & 1000  & \textbf{894} & \textbf{894.00 } & 0.00  & 6.8   &       & 894   & 893.40  & 110.9  &       & 0 \\
3dl141000 & 2744  & \textbf{2446} & \textbf{2445.80 } & 1.00  & 298.7  &       & 2434  & 2416.40  & 1039.7  &       & 12 \\
3dl142000 & 2744  & \textbf{2458} & \textbf{2458.00 } & 0.00  & 223.3  &       & 2444  & 2431.00  & 1016.2  &       & 14 \\
3dl143000 & 2744  & \textbf{2444} & \textbf{2440.60 } & 1.55  & 376.1  &       & 2426  & 2415.00  & 1012.3  &       & 18 \\
3dl144000 & 2744  & \textbf{2450} & \textbf{2448.20 } & 1.55  & 619.6  &       & 2440  & 2425.30  & 997.5  &       & 10 \\
3dl145000 & 2744  & \textbf{2446} & \textbf{2445.50 } & 1.61  & 475.1  &       & 2432  & 2422.40  & 999.3  &       & 14 \\
3dl146000 & 2744  & \textbf{2452} & \textbf{2450.50 } & 1.84  & 565.9  &       & 2438  & 2430.00  & 1035.4  &       & 14 \\
3dl147000 & 2744  & \textbf{2444} & \textbf{2442.10 } & 1.84  & 172.4  &       & 2428  & 2413.40  & 1022.7  &       & 16 \\
3dl148000 & 2744  & \textbf{2448} & \textbf{2446.10 } & 1.73  & 265.9  &       & 2432  & 2424.40  & 1030.7  &       & 16 \\
3dl149000 & 2744  & \textbf{2428} & \textbf{2425.20 } & 1.48  & 64.5  &       & 2418  & 2403.70  & 1020.1  &       & 10 \\
3dl1410000 & 2744  & \textbf{2458} & \textbf{2456.80 } & 2.00  & 538.2  &       & 2438  & 2429.30  & 1018.1  &       & 20 \\
\hline
    Better &            &   50/74/91 &   71/74/91 &            &            &            &            &            &            &            &            \\

     Equal &            &   24/74/91 &    3/74/91 &            &            &            &            &            &            &            &            \\

     Worse &            &    0/74/91 &    0/74/91 &            &            &            &            &            &            &            &            \\

\bottomrule
\end{longtable}
}

{\tiny
\begin{longtable}{@{}lrrrrrrrrrrr@{}}
\caption{Comparative results for max-3-cut between the proposed MOH algorithm and DC \cite{Zhu2013}} \label{table-comp-3cut}\\
\toprule
  Instance &     $|V|$ &    \multicolumn{ 4}{c}{MOH} &   & \multicolumn{ 2}{c}{DC} &  & $gap$\\
\cmidrule{3-6} \cmidrule{8-9}
  &   &   $f_{best}$&   $f_{avg}$&   $std$&   $time(s)$ &   & $f_{best}$ &  $time(s)$ &  &\\
\midrule
\endfirsthead

\multicolumn{12}{c}%
{{\bfseries \tablename\ \thetable{} -- continued from previous page}} \\
\toprule
    Instance &     $|V|$ &    \multicolumn{ 4}{c}{MOH} &   & \multicolumn{ 2}{c}{DC} &  & $gap$  \\
\cmidrule{3-6} \cmidrule{8-9}
  &   &   $f_{best}$&   $f_{avg}$&   $std$&   $time(s)$ &   & $f_{best}$ &    $time(s)$ &  &  \\
\midrule
\endhead
\midrule
\endfoot

\hline \hline
\endlastfoot

G1    & 800   & \textbf{15165} & 15164.90  & 0.36  & 605.4  &       & 15127 & 363.1  &       & 38 \\
G2    & 800   & \textbf{15172} & 15171.20  & 0.99  & 539.2  &       & 15159 & 355.4  &       & 13 \\
G3    & 800   & \textbf{15173} & 15173.00  & 0.00  & 227.4  &       & 15149 & 361.8  &       & 24 \\
G4    & 800   & 15184 & 15181.40  & 2.46  & 657.0  &       & -     & -     &       & - \\
G5    & 800   & 15193 & 15193.00  & 0.00  & 81.0  &       & -     & -     &       & - \\
G6    & 800   & 2632  & 2631.95  & 0.22  & 269.6  &       & -     & -     &       & - \\
G7    & 800   & 2409  & 2408.40  & 1.07  & 491.3  &       & -     & -     &       & - \\
G8    & 800   & 2428  & 2427.55  & 0.67  & 682.5  &       & -     & -     &       & - \\
G9    & 800   & 2478  & 2475.85  & 2.52  & 692.4  &       & -     & -     &       & - \\
G10   & 800   & 2407  & 2406.40  & 0.86  & 930.9  &       & -     & -     &       & - \\
G11   & 800   & \textbf{669} & 667.80  & 0.75  & 708.9  &       & 660   & 172.1  &       & 9 \\
G12   & 800   & \textbf{660} & 658.95  & 0.50  & 992.9  &       & 655   & 151.8  &       & 5 \\
G13   & 800   & \textbf{686} & 685.40  & 0.58  & 586.8  &       & 679   & 164.4  &       & 7 \\
G14   & 800   & \textbf{4012} & 4009.45  & 1.88  & 45.7  &       & 3984  & 193.9  &       & 28 \\
G15   & 800   & \textbf{3984} & 3982.40  & 0.58  & 282.0  &       & 3960  & 194.2  &       & 24 \\
G16   & 800   & \textbf{3991} & 3986.30  & 1.87  & 10.8  &       & 3958  & 194.6  &       & 33 \\
G17   & 800   & 3983  & 3981.00  & 1.05  & 79.9  &       & -     & -     &       & - \\
G18   & 800   & 1207  & 1205.60  & 1.56  & 5.9   &       & -     & -     &       & - \\
G19   & 800   & 1081  & 1078.05  & 2.38  & 3.0   &       & -     & -     &       & - \\
G20   & 800   & 1122  & 1115.00  & 4.05  & 16.1  &       & -     & -     &       & - \\
G21   & 800   & 1109  & 1106.75  & 2.30  & 90.9  &       & -     & -     &       & - \\
G22   & 2000  & \textbf{17167} & 17157.80  & 7.62  & 561.0  &       & 17008 & 1515.3  &       & 159 \\
G23   & 2000  & \textbf{17168} & 17156.70  & 6.40  & 888.4  &       & 17021 & 1564.5  &       & 147 \\
G24   & 2000  & \textbf{17162} & 17152.10  & 4.98  & 321.4  &       & 17037 & 1592.9  &       & 125 \\
G25   & 2000  & 17163 & 17155.20  & 3.44  & 1276.8  &       & -     & -     &       & - \\
G26   & 2000  & 17154 & 17146.30  & 4.61  & 883.4  &       & -     & -     &       & - \\
G27   & 2000  & 4020  & 4013.80  & 3.33  & 576.8  &       & -     & -     &       & - \\
G28   & 2000  & 3973  & 3966.45  & 5.10  & 766.1  &       & -     & -     &       & - \\
G29   & 2000  & 4106  & 4097.30  & 5.40  & 285.6  &       & -     & -     &       & - \\
G30   & 2000  & 4119  & 4109.90  & 5.34  & 1482.9  &       & -     & -     &       & - \\
G31   & 2000  & 4003  & 3999.20  & 6.69  & 819.7  &       & -     & -     &       & - \\
G32   & 2000  & \textbf{1653} & 1651.85  & 0.73  & 522.3  &       & 1635  & 910.7  &       & 18 \\
G33   & 2000  & \textbf{1625} & 1622.30  & 0.95  & 1233.4  &       & 1603  & 868.0  &       & 22 \\
G34   & 2000  & \textbf{1607} & 1604.00  & 1.00  & 1752.1  &       & 1589  & 931.3  &       & 18 \\
G35   & 2000  & \textbf{10046} & 10039.90  & 2.59  & 1304.4  &       & 9965  & 1280.9  &       & 81 \\
G36   & 2000  & \textbf{10039} & 10034.40  & 3.81  & 1291.6  &       & 9945  & 1301.5  &       & 94 \\
G37   & 2000  & \textbf{10052} & 10047.80  & 1.96  & 64.1  &       & 9952  & 1318.0  &       & 100 \\
G38   & 2000  & 10040 & 10035.50  & 3.26  & 888.4  &       & -     & -     &       & - \\
G39   & 2000  & 2903  & 2890.05  & 6.75  & 176.5  &       & -     & -     &       & - \\
G40   & 2000  & 2870  & 2850.65  & 8.08  & 1632.8  &       & -     & -     &       & - \\
G41   & 2000  & 2887  & 2862.90  & 9.77  & 1729.4  &       & -     & -     &       & - \\
G42   & 2000  & 2980  & 2964.30  & 5.99  & 48.3  &       & -     & -     &       & - \\
G43   & 1000  & \textbf{8573} & 8573.00  & 0.00  & 282.2  &       & 8510  & 366.1  &       & 63 \\
G44   & 1000  & \textbf{8571} & 8569.60  & 2.35  & 705.5  &       & 8526  & 351.0  &       & 45 \\
G45   & 1000  & \textbf{8566} & 8564.85  & 1.11  & 246.5  &       & 8515  & 360.1  &       & 51 \\
G46   & 1000  & 8568  & 8564.60  & 2.01  & 1061.4  &       & -     & -     &       & - \\
G47   & 1000  & 8572  & 8568.70  & 2.72  & 621.5  &       & -     & -     &       & - \\
G48   & 3000  & \textbf{6000} & 6000.00  & 0.00  & 0.3   &       & 5998  & 1850.9  &       & 2 \\
G49   & 3000  & 6000  & 6000.00  & 0.00  & 0.7   &       & 6000  & 1895.3  &       & 0 \\
G50   & 3000  & \textbf{6000} & 6000.00  & 0.00  & 116.5  &       & 5998  & 1819.8  &       & 2 \\
G51   & 1000  & 5037  & 5031.35  & 1.90  & 944.6  &       & -     & -     &       & - \\
G52   & 1000  & 5040  & 5037.50  & 0.81  & 12.8  &       & -     & -     &       & - \\
G53   & 1000  & 5039  & 5038.00  & 1.05  & 307.2  &       & -     & -     &       & - \\
G54   & 1000  & 5036  & 5033.55  & 2.29  & 880.1  &       & -     & -     &       & - \\
G55   & 5000  & 12429 & 12423.70  & 2.61  & 6573.0  &       & -     & -     &       & - \\
G56   & 5000  & 4752  & 4741.90  & 7.84  & 1168.4  &       & -     & -     &       & - \\
G57   & 5000  & 4083  & 4079.00  & 1.55  & 5457.3  &       & -     & -     &       & - \\
G58   & 5000  & 25195 & 25182.10  & 8.89  & 397.3  &       & -     & -     &       & - \\
G59   & 5000  & 7262  & 7246.70  & 9.20  & 3575.1  &       & -     & -     &       & - \\
G60   & 7000  & 17076 & 17067.00  & 4.40  & 6745.0  &       & -     & -     &       & - \\
G61   & 7000  & 6853  & 6842.10  & 5.26  & 3608.6  &       & -     & -     &       & - \\
G62   & 7000  & 5685  & 5681.50  & 1.43  & 6250.1  &       & -     & -     &       & - \\
G63   & 7000  & 35322 & 35301.60  & 10.35  & 6546.8  &       & -     & -     &       & - \\
G64   & 7000  & 10443 & 10408.80  & 25.23  & 1563.7  &       & -     & -     &       & - \\
G65   & 8000  & 6490  & 6485.80  & 2.04  & 3077.6  &       & -     & -     &       & - \\
G66   & 9000  & 7416  & 7411.50  & 2.42  & 5126.0  &       & -     & -     &       & - \\
G67   & 10000 & 8086  & 8083.50  & 2.29  & 1048.1  &       & -     & -     &       & - \\
G70   & 10000 & 9999  & 9999.00  & 0.00  & 5.6   &       & -     & -     &       & - \\
G72   & 10000 & 8192  & 8186.70  & 3.35  & 6393.0  &       & -     & -     &       & - \\
G77   & 14000 & 11578 & 11568.90  & 4.01  & 1899.0  &       & -     & -     &       & - \\
G81   & 20000 & 16321 & 16313.00  & 4.05  & 4821.4  &       & -     & -     &       & - \\
3dl101000 & 1000  & \textbf{1067} & 1066.10  & 0.54  & 679.6  &       & 1043  & 238.2  &       & 24 \\
3dl102000 & 1000  & \textbf{1072} & 1071.95  & 0.22  & 560.9  &       & 1044  & 242.4  &       & 28 \\
3dl103000 & 1000  & \textbf{1065} & 1063.60  & 0.66  & 1303.4  &       & 1042  & 233.4  &       & 23 \\
3dl104000 & 1000  & \textbf{1071} & 1070.30  & 0.46  & 526.5  &       & 1045  & 244.0  &       & 26 \\
3dl105000 & 1000  & \textbf{1064} & 1061.90  & 0.77  & 71.0  &       & 1039  & 229.2  &       & 25 \\
3dl106000 & 1000  & \textbf{1063} & 1061.80  & 0.60  & 882.4  &       & 1032  & 252.7  &       & 31 \\
3dl107000 & 1000  & \textbf{1075} & 1074.40  & 0.58  & 467.2  &       & 1053  & 240.0  &       & 22 \\
3dl108000 & 1000  & \textbf{1071} & 1069.95  & 0.38  & 178.5  &       & 1049  & 232.5  &       & 22 \\
3dl109000 & 1000  & \textbf{1079} & 1078.20  & 0.81  & 510.1  &       & 1052  & 234.6  &       & 27 \\
3dl1010000 & 1000  & \textbf{1070} & 1069.50  & 0.50  & 493.8  &       & 1044  & 247.2  &       & 26 \\
3dl141000 & 2744  & \textbf{2924} & 2919.75  & 2.45  & 493.0  &       & 2845  & 1805.5  &       & 79 \\
3dl142000 & 2744  & \textbf{2935} & 2929.25  & 2.53  & 1103.3  &       & 2856  & 1826.3  &       & 79 \\
3dl143000 & 2744  & \textbf{2912} & 2909.50  & 1.40  & 1087.0  &       & 2829  & 1898.8  &       & 83 \\
3dl144000 & 2744  & \textbf{2924} & 2919.90  & 2.41  & 458.5  &       & 2861  & 1779.2  &       & 63 \\
3dl145000 & 2744  & \textbf{2914} & 2911.25  & 1.92  & 665.5  &       & 2839  & 1796.7  &       & 75 \\
3dl146000 & 2744  & \textbf{2913} & 2909.00  & 2.00  & 331.3  &       & 2834  & 1815.3  &       & 79 \\
3dl147000 & 2744  & \textbf{2913} & 2909.30  & 1.73  & 1381.3  &       & 2834  & 1824.4  &       & 79 \\
3dl148000 & 2744  & \textbf{2925} & 2919.40  & 4.05  & 729.1  &       & 2845  & 1782.1  &       & 80 \\
3dl149000 & 2744  & \textbf{2906} & 2901.50  & 2.62  & 125.2  &       & 2823  & 1768.9  &       & 83 \\
3dl1410000 & 2744  & \textbf{2933} & 2927.65  & 2.22  & 589.6  &       & 2851  & 1799.4  &       & 82 \\
\hline
    Better &            &   43/44/91 &            &            &            &            &            &            &            &            \\

     Equal &            &    1/44/91 &            &            &            &            &            &            &            &            \\

     Worse &            &    0/44/91 &            &            &            &            &            &            &            &            \\

\bottomrule
\end{longtable}
}


{\tiny
\begin{longtable}{@{}lrrrrrrrrrrr@{}}
\caption{Comparative results for max-4-cut between the proposed MOH algorithm and DC \cite{Zhu2013}}\label{table-comp-4cut}\\
\toprule
  Instance &     $|V|$ &    \multicolumn{ 4}{c}{MOH} &   & \multicolumn{ 2}{c}{DC} &  & $gap$\\
\cmidrule{3-6} \cmidrule{8-9}
  &   &   $f_{best}$&   $f_{avg}$&   $std$&   $time(s)$ &   & $f_{best}$ &  $time(s)$ &  &\\
\midrule
\endfirsthead
\multicolumn{12}{c}%
{{\bfseries \tablename\ \thetable{} -- continued from previous page}} \\
\toprule
    Instance &     $|V|$ &    \multicolumn{ 4}{c}{MOH} &   & \multicolumn{ 2}{c}{DC} &  & $gap$  \\
\cmidrule{3-6} \cmidrule{8-9}

  &   &   $f_{best}$&   $f_{avg}$&   $std$&   $time(s)$ &   & $f_{best}$ &    $time(s)$ &  &  \\
\midrule
\endhead
\midrule
\endfoot
\hline \hline
\endlastfoot
G1    & 800   & \textbf{16803} & 16801.00  & 0.87  & 522.1  &       & 16740 & 450.2  &       & 63 \\
G2    & 800   & \textbf{16809} & 16808.00  & 1.12  & 694.2  &       & 16735 & 455.8  &       & 74 \\
G3    & 800   & \textbf{16806} & 16804.70  & 1.05  & 909.6  &       & 16752 & 431.9  &       & 54 \\
G4    & 800   & 16814 & 16811.20  & 1.50  & 967.7  &       & -     & -     &       & - \\
G5    & 800   & 16816 & 16815.80  & 0.92  & 628.0  &       & -     & -     &       & - \\
G6    & 800   & 2751  & 2748.45  & 1.16  & 1775.5  &       & -     & -     &       & - \\
G7    & 800   & 2515  & 2513.75  & 0.92  & 1128.1  &       & -     & -     &       & - \\
G8    & 800   & 2525  & 2523.35  & 0.74  & 1551.5  &       & -     & -     &       & - \\
G9    & 800   & 2585  & 2583.35  & 1.02  & 324.7  &       & -     & -     &       & - \\
G10   & 800   & 2510  & 2507.60  & 1.38  & 788.1  &       & -     & -     &       & - \\
G11   & 800   & \textbf{677} & 676.00  & 0.32  & 400.7  &       & 675   & 171.3  &       & 2 \\
G12   & 800   & \textbf{664} & 662.25  & 0.59  & 814.2  &       & 660   & 180.0  &       & 4 \\
G13   & 800   & \textbf{690} & 689.10  & 0.45  & 689.2  &       & 685   & 187.5  &       & 5 \\
G14   & 800   & \textbf{4440} & 4435.35  & 1.96  & 1095.5  &       & 4402  & 243.1  &       & 38 \\
G15   & 800   & \textbf{4406} & 4403.40  & 0.89  & 1757.7  &       & 4373  & 249.7  &       & 33 \\
G16   & 800   & \textbf{4415} & 4414.05  & 1.02  & 957.2  &       & 4378  & 246.1  &       & 37 \\
G17   & 800   & 4411  & 4406.45  & 2.31  & 3.9   &       & -     & -     &       & - \\
G18   & 800   & 1261  & 1253.90  & 3.19  & 5.9   &       & -     & -     &       & - \\
G19   & 800   & 1121  & 1115.35  & 3.71  & 6.6   &       & -     & -     &       & - \\
G20   & 800   & 1168  & 1160.95  & 3.26  & 7.9   &       & -     & -     &       & - \\
G21   & 800   & 1155  & 1148.25  & 3.75  & 1079.7  &       & -     & -     &       & - \\
G22   & 2000  & \textbf{18776} & 18765.70  & 5.71  & 1013.6  &       & 18615 & 1988.3  &       & 161 \\
G23   & 2000  & \textbf{18777} & 18765.80  & 5.77  & 1454.7  &       & 18612 & 1941.9  &       & 165 \\
G24   & 2000  & \textbf{18769} & 18763.60  & 3.79  & 521.1  &       & 18620 & 1822.8  &       & 149 \\
G25   & 2000  & 18775 & 18767.60  & 4.40  & 1493.2  &       & -     & -     &       & - \\
G26   & 2000  & 18767 & 18761.20  & 4.49  & 635.3  &       & -     & -     &       & - \\
G27   & 2000  & 4201  & 4188.50  & 4.63  & 754.0  &       & -     & -     &       & - \\
G28   & 2000  & 4150  & 4138.85  & 5.97  & 492.5  &       & -     & -     &       & - \\
G29   & 2000  & 4293  & 4281.65  & 5.71  & 1725.1  &       & -     & -     &       & - \\
G30   & 2000  & 4305  & 4296.40  & 4.14  & 661.2  &       & -     & -     &       & - \\
G31   & 2000  & 4171  & 4164.40  & 6.47  & 1063.9  &       & -     & -     &       & - \\
G32   & 2000  & \textbf{1669} & 1667.85  & 1.32  & 349.0  &       & 1659  & 1140.7  &       & 10 \\
G33   & 2000  & \textbf{1638} & 1634.65  & 1.32  & 0.0   &       & 1629  & 1052.4  &       & 9 \\
G34   & 2000  & \textbf{1616} & 1611.70  & 1.79  & 1.0   &       & 1604  & 1105.0  &       & 12 \\
G35   & 2000  & \textbf{11111} & 11106.20  & 2.16  & 324.7  &       & 11007 & 1890.3  &       & 104 \\
G36   & 2000  & \textbf{11108} & 11101.40  & 2.92  & 340.5  &       & 10993 & 1738.6  &       & 115 \\
G37   & 2000  & \textbf{11117} & 11112.50  & 2.40  & 693.8  &       & 11023 & 1754.2  &       & 94 \\
G38   & 2000  & 11108 & 11101.10  & 3.16  & 955.3  &       & -     & -     &       & - \\
G39   & 2000  & 3006  & 2998.70  & 3.97  & 22.7  &       & -     & -     &       & - \\
G40   & 2000  & 2976  & 2955.65  & 9.01  & 961.3  &       & -     & -     &       & - \\
G41   & 2000  & 2983  & 2970.30  & 6.91  & 35.5  &       & -     & -     &       & - \\
G42   & 2000  & 3092  & 3084.05  & 4.80  & 285.2  &       & -     & -     &       & - \\
G43   & 1000  & \textbf{9376} & 9373.95  & 1.53  & 1656.1  &       & 9306  & 423.0  &       & 70 \\
G44   & 1000  & \textbf{9379} & 9373.55  & 2.58  & 1340.2  &       & 9315  & 430.5  &       & 64 \\
G45   & 1000  & \textbf{9376} & 9375.10  & 0.95  & 612.1  &       & 9312  & 463.5  &       & 64 \\
G46   & 1000  & 9378  & 9375.35  & 1.99  & 639.0  &       & -     & -     &       & - \\
G47   & 1000  & 9381  & 9377.05  & 2.04  & 1194.2  &       & -     & -     &       & - \\
G48   & 3000  & 6000  & 6000.00  & 0.00  & 0.0   &       & 6000  & 1673.8  &       & 0 \\
G49   & 3000  & 6000  & 6000.00  & 0.00  & 0.0   &       & 6000  & 1675.6  &       & 0 \\
G50   & 3000  & 6000  & 6000.00  & 0.00  & 0.0   &       & 6000  & 1678.9  &       & 0 \\
G51   & 1000  & 5571  & 5567.65  & 2.04  & 143.6  &       & -     & -     &       & - \\
G52   & 1000  & 5584  & 5581.15  & 1.75  & 129.9  &       & -     & -     &       & - \\
G53   & 1000  & 5574  & 5571.85  & 1.47  & 67.1  &       & -     & -     &       & - \\
G54   & 1000  & 5579  & 5576.25  & 1.60  & 13.8  &       & -     & -     &       & - \\
G55   & 5000  & 12498 & 12498.00  & 0.00  & 0.1   &       & -     & -     &       & - \\
G56   & 5000  & 4931  & 4917.10  & 6.49  & 4190.5  &       & -     & -     &       & - \\
G57   & 5000  & 4112  & 4110.50  & 1.22  & 2942.0  &       & -     & -     &       & - \\
G58   & 5000  & 27885 & 27870.90  & 8.72  & 4297.1  &       & -     & -     &       & - \\
G59   & 5000  & 7539  & 7515.10  & 15.09  & 4782.7  &       & -     & -     &       & - \\
G60   & 7000  & 17148 & 17148.00  & 0.00  & 1.4   &       & -     & -     &       & - \\
G61   & 7000  & 7110  & 7104.60  & 5.12  & 6440.2  &       & -     & -     &       & - \\
G62   & 7000  & 5743  & 5738.70  & 2.77  & 3804.6  &       & -     & -     &       & - \\
G63   & 7000  & 39083 & 39063.50  & 9.19  & 6515.7  &       & -     & -     &       & - \\
G64   & 7000  & 10814 & 10797.40  & 13.29  & 4493.0  &       & -     & -     &       & - \\
G65   & 8000  & 6534  & 6525.40  & 4.49  & 14.8  &       & -     & -     &       & - \\
G66   & 9000  & 7474  & 7467.80  & 4.31  & 21.7  &       & -     & -     &       & - \\
G67   & 10000 & 8155  & 8142.50  & 5.59  & 29.6  &       & -     & -     &       & - \\
G70   & 10000 & 9999  & 9999.00  & 0.00  & 0.2   &       & -     & -     &       & - \\
G72   & 10000 & 8264  & 8254.60  & 7.39  & 15.3  &       & -     & -     &       & - \\
G77   & 14000 & 11674 & 11658.90  & 10.12  & 63.2  &       & -     & -     &       & - \\
G81   & 20000 & 16470 & 16454.30  & 8.50  & 271.4  &       & -     & -     &       & - \\
3dl101000 & 1000  & \textbf{1103} & 1100.60  & 1.05  & 1273.1  &       & 1073  & 304.4  &       & 30 \\
3dl102000 & 1000  & \textbf{1102} & 1100.00  & 0.95  & 29.6  &       & 1070  & 351.3  &       & 32 \\
3dl103000 & 1000  & \textbf{1108} & 1106.40  & 0.95  & 225.0  &       & 1072  & 341.0  &       & 36 \\
3dl104000 & 1000  & \textbf{1103} & 1101.65  & 0.92  & 564.5  &       & 1076  & 323.5  &       & 27 \\
3dl105000 & 1000  & \textbf{1098} & 1096.30  & 0.84  & 578.3  &       & 1074  & 334.4  &       & 24 \\
3dl106000 & 1000  & \textbf{1097} & 1095.15  & 0.92  & 928.2  &       & 1063  & 358.3  &       & 34 \\
3dl107000 & 1000  & \textbf{1114} & 1112.20  & 1.10  & 712.6  &       & 1093  & 308.3  &       & 21 \\
3dl108000 & 1000  & \textbf{1105} & 1103.00  & 0.77  & 478.7  &       & 1079  & 276.1  &       & 26 \\
3dl109000 & 1000  & \textbf{1115} & 1113.45  & 0.92  & 641.0  &       & 1086  & 271.3  &       & 29 \\
3dl1010000 & 1000  & \textbf{1109} & 1106.10  & 0.89  & 1083.6  &       & 1088  & 277.2  &       & 21 \\
3dl141000 & 2744  & \textbf{3016} & 3012.05  & 1.91  & 563.0  &       & 2893  & 1990.5  &       & 123 \\
3dl142000 & 2744  & \textbf{3026} & 3019.80  & 2.19  & 364.2  &       & 2893  & 2007.3  &       & 133 \\
3dl143000 & 2744  & \textbf{3006} & 3001.70  & 2.97  & 367.1  &       & 2892  & 1956.1  &       & 114 \\
3dl144000 & 2744  & \textbf{3012} & 3007.85  & 2.04  & 943.5  &       & 2897  & 1980.3  &       & 115 \\
3dl145000 & 2744  & \textbf{3006} & 3001.20  & 2.17  & 1146.8  &       & 2882  & 1972.2  &       & 124 \\
3dl146000 & 2744  & \textbf{3005} & 3001.35  & 1.50  & 256.6  &       & 2888  & 1948.9  &       & 117 \\
3dl147000 & 2744  & \textbf{3007} & 3001.95  & 2.50  & 301.0  &       & 2879  & 1995.7  &       & 128 \\
3dl148000 & 2744  & \textbf{3018} & 3014.50  & 2.02  & 1632.9  &       & 2883  & 1982.7  &       & 135 \\
3dl149000 & 2744  & \textbf{2999} & 2993.95  & 2.78  & 394.8  &       & 2877  & 2024.5  &       & 122 \\
3dl1410000 & 2744  & \textbf{3023} & 3021.15  & 1.69  & 1075.8  &       & 2904  & 2007.4  &       & 119 \\
\hline
    Better &            &   41/44/91 &            &            &            &            &            &            &            &            \\

     Equal &            &    3/44/91 &            &            &            &            &            &            &            &            \\

     Worse &            &    0/44/91 &            &            &            &            &            &            &            &            \\

\bottomrule
\end{longtable}
}

{\tiny
\begin{longtable}{@{}lrrrrrrrrrrr@{}}
\caption{Comparative results for max-5-cut between the proposed MOH algorithm and DC \cite{Zhu2013}}\label{table-comp-5cut}\\

\toprule
  Instance &     $|V|$ &    \multicolumn{ 4}{c}{MOH} &   & \multicolumn{ 2}{c}{DC} &  & $gap$\\
\cmidrule{3-6} \cmidrule{8-9}

  &   &   $f_{best}$&   $f_{avg}$&   $std$&   $time(s)$ &   & $f_{best}$ &  $time(s)$ &  &\\
\midrule
\endfirsthead
\multicolumn{12}{c}%
{{\bfseries \tablename\ \thetable{} -- continued from previous page}} \\
\toprule
    Instance &     $|V|$ &    \multicolumn{ 4}{c}{MOH} &   & \multicolumn{ 2}{c}{DC} &  & $gap$  \\
\cmidrule{3-6} \cmidrule{8-9}

  &   &   $f_{best}$&   $f_{avg}$&   $std$&   $time(s)$ &   & $f_{best}$ &    $time(s)$ &  &  \\
\midrule
\endhead
\midrule
\endfoot

\hline \hline
\endlastfoot

G1    & 800   & \textbf{16803} & 16801.00  & 0.87  & 522.1  &       & 16740 & 450.2  &       & 63 \\
G2    & 800   & \textbf{16809} & 16808.00  & 1.12  & 694.2  &       & 16735 & 455.8  &       & 74 \\
G3    & 800   & \textbf{16806} & 16804.70  & 1.05  & 909.6  &       & 16752 & 431.9  &       & 54 \\
G4    & 800   & 16814 & 16811.20  & 1.50  & 967.7  &       & -     & -     &       & - \\
G5    & 800   & 16816 & 16815.80  & 0.92  & 628.0  &       & -     & -     &       & - \\
G6    & 800   & 2751  & 2748.45  & 1.16  & 1775.5  &       & -     & -     &       & - \\
G7    & 800   & 2515  & 2513.75  & 0.92  & 1128.1  &       & -     & -     &       & - \\
G8    & 800   & 2525  & 2523.35  & 0.74  & 1551.5  &       & -     & -     &       & - \\
G9    & 800   & 2585  & 2583.35  & 1.02  & 324.7  &       & -     & -     &       & - \\
G10   & 800   & 2510  & 2507.60  & 1.38  & 788.1  &       & -     & -     &       & - \\
G11   & 800   & \textbf{677} & 676.00  & 0.32  & 400.7  &       & 675   & 171.3  &       & 2 \\
G12   & 800   & \textbf{664} & 662.25  & 0.59  & 814.2  &       & 660   & 180.0  &       & 4 \\
G13   & 800   & \textbf{690} & 689.10  & 0.45  & 689.2  &       & 685   & 187.5  &       & 5 \\
G14   & 800   & \textbf{4440} & 4435.35  & 1.96  & 1095.5  &       & 4402  & 243.1  &       & 38 \\
G15   & 800   & \textbf{4406} & 4403.40  & 0.89  & 1757.7  &       & 4373  & 249.7  &       & 33 \\
G16   & 800   & \textbf{4415} & 4414.05  & 1.02  & 957.2  &       & 4378  & 246.1  &       & 37 \\
G17   & 800   & 4411  & 4406.45  & 2.31  & 3.9   &       & -     & -     &       & - \\
G18   & 800   & 1261  & 1253.90  & 3.19  & 5.9   &       & -     & -     &       & - \\
G19   & 800   & 1121  & 1115.35  & 3.71  & 6.6   &       & -     & -     &       & - \\
G20   & 800   & 1168  & 1160.95  & 3.26  & 7.9   &       & -     & -     &       & - \\
G21   & 800   & 1155  & 1148.25  & 3.75  & 1079.7  &       & -     & -     &       & - \\
G22   & 2000  & \textbf{18776} & 18765.70  & 5.71  & 1013.6  &       & 18615 & 1988.3  &       & 161 \\
G23   & 2000  & \textbf{18777} & 18765.80  & 5.77  & 1454.7  &       & 18612 & 1941.9  &       & 165 \\
G24   & 2000  & \textbf{18769} & 18763.60  & 3.79  & 521.1  &       & 18620 & 1822.8  &       & 149 \\
G25   & 2000  & 18775 & 18767.60  & 4.40  & 1493.2  &       & -     & -     &       & - \\
G26   & 2000  & 18767 & 18761.20  & 4.49  & 635.3  &       & -     & -     &       & - \\
G27   & 2000  & 4201  & 4188.50  & 4.63  & 754.0  &       & -     & -     &       & - \\
G28   & 2000  & 4150  & 4138.85  & 5.97  & 492.5  &       & -     & -     &       & - \\
G29   & 2000  & 4293  & 4281.65  & 5.71  & 1725.1  &       & -     & -     &       & - \\
G30   & 2000  & 4305  & 4296.40  & 4.14  & 661.2  &       & -     & -     &       & - \\
G31   & 2000  & 4171  & 4164.40  & 6.47  & 1063.9  &       & -     & -     &       & - \\
G32   & 2000  & \textbf{1669} & 1667.85  & 1.32  & 349.0  &       & 1659  & 1140.7  &       & 10 \\
G33   & 2000  & \textbf{1638} & 1634.65  & 1.32  & 0.0   &       & 1629  & 1052.4  &       & 9 \\
G34   & 2000  & \textbf{1616} & 1611.70  & 1.79  & 1.0   &       & 1604  & 1105.0  &       & 12 \\
G35   & 2000  & \textbf{11111} & 11106.20  & 2.16  & 324.7  &       & 11007 & 1890.3  &       & 104 \\
G36   & 2000  & \textbf{11108} & 11101.40  & 2.92  & 340.5  &       & 10993 & 1738.6  &       & 115 \\
G37   & 2000  & \textbf{11117} & 11112.50  & 2.40  & 693.8  &       & 11023 & 1754.2  &       & 94 \\
G38   & 2000  & 11108 & 11101.10  & 3.16  & 955.3  &       & -     & -     &       & - \\
G39   & 2000  & 3006  & 2998.70  & 3.97  & 22.7  &       & -     & -     &       & - \\
G40   & 2000  & 2976  & 2955.65  & 9.01  & 961.3  &       & -     & -     &       & - \\
G41   & 2000  & 2983  & 2970.30  & 6.91  & 35.5  &       & -     & -     &       & - \\
G42   & 2000  & 3092  & 3084.05  & 4.80  & 285.2  &       & -     & -     &       & - \\
G43   & 1000  & \textbf{9376} & 9373.95  & 1.53  & 1656.1  &       & 9306  & 423.0  &       & 70 \\
G44   & 1000  & \textbf{9379} & 9373.55  & 2.58  & 1340.2  &       & 9315  & 430.5  &       & 64 \\
G45   & 1000  & \textbf{9376} & 9375.10  & 0.95  & 612.1  &       & 9312  & 463.5  &       & 64 \\
G46   & 1000  & 9378  & 9375.35  & 1.99  & 639.0  &       & -     & -     &       & - \\
G47   & 1000  & 9381  & 9377.05  & 2.04  & 1194.2  &       & -     & -     &       & - \\
G48   & 3000  & 6000  & 6000.00  & 0.00  & 0.0   &       & 6000  & 1673.8  &       & 0 \\
G49   & 3000  & 6000  & 6000.00  & 0.00  & 0.0   &       & 6000  & 1675.6  &       & 0 \\
G50   & 3000  & 6000  & 6000.00  & 0.00  & 0.0   &       & 6000  & 1678.9  &       & 0 \\
G51   & 1000  & 5571  & 5567.65  & 2.04  & 143.6  &       & -     & -     &       & - \\
G52   & 1000  & 5584  & 5581.15  & 1.75  & 129.9  &       & -     & -     &       & - \\
G53   & 1000  & 5574  & 5571.85  & 1.47  & 67.1  &       & -     & -     &       & - \\
G54   & 1000  & 5579  & 5576.25  & 1.60  & 13.8  &       & -     & -     &       & - \\
G55   & 5000  & 12498 & 12498.00  & 0.00  & 0.1   &       & -     & -     &       & - \\
G56   & 5000  & 4931  & 4917.10  & 6.49  & 4190.5  &       & -     & -     &       & - \\
G57   & 5000  & 4112  & 4110.50  & 1.22  & 2942.0  &       & -     & -     &       & - \\
G58   & 5000  & 27885 & 27870.90  & 8.72  & 4297.1  &       & -     & -     &       & - \\
G59   & 5000  & 7539  & 7515.10  & 15.09  & 4782.7  &       & -     & -     &       & - \\
G60   & 7000  & 17148 & 17148.00  & 0.00  & 1.4   &       & -     & -     &       & - \\
G61   & 7000  & 7110  & 7104.60  & 5.12  & 6440.2  &       & -     & -     &       & - \\
G62   & 7000  & 5743  & 5738.70  & 2.77  & 3804.6  &       & -     & -     &       & - \\
G63   & 7000  & 39083 & 39063.50  & 9.19  & 6515.7  &       & -     & -     &       & - \\
G64   & 7000  & 10814 & 10797.40  & 13.29  & 4493.0  &       & -     & -     &       & - \\
G65   & 8000  & 6534  & 6525.40  & 4.49  & 14.8  &       & -     & -     &       & - \\
G66   & 9000  & 7474  & 7467.80  & 4.31  & 21.7  &       & -     & -     &       & - \\
G67   & 10000 & 8155  & 8142.50  & 5.59  & 29.6  &       & -     & -     &       & - \\
G70   & 10000 & 9999  & 9999.00  & 0.00  & 0.2   &       & -     & -     &       & - \\
G72   & 10000 & 8264  & 8254.60  & 7.39  & 15.3  &       & -     & -     &       & - \\
G77   & 14000 & 11674 & 11658.90  & 10.12  & 63.2  &       & -     & -     &       & - \\
G81   & 20000 & 16470 & 16454.30  & 8.50  & 271.4  &       & -     & -     &       & - \\
3dl101000 & 1000  & \textbf{1103} & 1100.60  & 1.05  & 1273.1  &       & 1073  & 304.4  &       & 30 \\
3dl102000 & 1000  & \textbf{1102} & 1100.00  & 0.95  & 29.6  &       & 1070  & 351.3  &       & 32 \\
3dl103000 & 1000  & \textbf{1108} & 1106.40  & 0.95  & 225.0  &       & 1072  & 341.0  &       & 36 \\
3dl104000 & 1000  & \textbf{1103} & 1101.65  & 0.92  & 564.5  &       & 1076  & 323.5  &       & 27 \\
3dl105000 & 1000  & \textbf{1098} & 1096.30  & 0.84  & 578.3  &       & 1074  & 334.4  &       & 24 \\
3dl106000 & 1000  & \textbf{1097} & 1095.15  & 0.92  & 928.2  &       & 1063  & 358.3  &       & 34 \\
3dl107000 & 1000  & \textbf{1114} & 1112.20  & 1.10  & 712.6  &       & 1093  & 308.3  &       & 21 \\
3dl108000 & 1000  & \textbf{1105} & 1103.00  & 0.77  & 478.7  &       & 1079  & 276.1  &       & 26 \\
3dl109000 & 1000  & \textbf{1115} & 1113.45  & 0.92  & 641.0  &       & 1086  & 271.3  &       & 29 \\
3dl1010000 & 1000  & \textbf{1109} & 1106.10  & 0.89  & 1083.6  &       & 1088  & 277.2  &       & 21 \\
3dl141000 & 2744  & \textbf{3016} & 3012.05  & 1.91  & 563.0  &       & 2893  & 1990.5  &       & 123 \\
3dl142000 & 2744  & \textbf{3026} & 3019.80  & 2.19  & 364.2  &       & 2893  & 2007.3  &       & 133 \\
3dl143000 & 2744  & \textbf{3006} & 3001.70  & 2.97  & 367.1  &       & 2892  & 1956.1  &       & 114 \\
3dl144000 & 2744  & \textbf{3012} & 3007.85  & 2.04  & 943.5  &       & 2897  & 1980.3  &       & 115 \\
3dl145000 & 2744  & \textbf{3006} & 3001.20  & 2.17  & 1146.8  &       & 2882  & 1972.2  &       & 124 \\
3dl146000 & 2744  & \textbf{3005} & 3001.35  & 1.50  & 256.6  &       & 2888  & 1948.9  &       & 117 \\
3dl147000 & 2744  & \textbf{3007} & 3001.95  & 2.50  & 301.0  &       & 2879  & 1995.7  &       & 128 \\
3dl148000 & 2744  & \textbf{3018} & 3014.50  & 2.02  & 1632.9  &       & 2883  & 1982.7  &       & 135 \\
3dl149000 & 2744  & \textbf{2999} & 2993.95  & 2.78  & 394.8  &       & 2877  & 2024.5  &       & 122 \\
3dl1410000 & 2744  & \textbf{3023} & 3021.15  & 1.69  & 1075.8  &       & 2904  & 2007.4  &       & 119 \\
\hline
    Better &            &   41/44/91 &            &            &            &            &            &            &            &            \\

     Equal &            &    3/44/91 &            &            &            &            &            &            &            &            \\

     Worse &            &    0/44/91 &            &            &            &            &            &            &            &            \\

\bottomrule
\end{longtable}
}

{\tiny
\renewcommand{\tabcolsep}{0.1cm}
\begin{longtable}{@{}llrrrrrrrr@{}}
\caption{Comparative results of the proposed MOH algorithm with 6 state of the art max-cut algorithms}\label{table-comp-5ref}\\
\toprule
  Instance & $|V|$ & $f_{pre}$ & GES \cite{shylo2012solving} & BLS \cite{benlic2013breakout} & MACUT \cite{wu2012memetic} &    TS-UBQP \cite{kochenberger2013solving} & TS/PM \cite{wang2013probabilistic}&  MAMBP \cite{wu2013memetic} & MOH \\
\midrule
\endfirsthead
\multicolumn{10}{c}%
{{\bfseries \tablename\ \thetable{} -- continued from previous page}} \\
\toprule
   Instance &     $|V|$ & $f_{pre}$ & GES \cite{shylo2012solving} & BLS \cite{benlic2013breakout} & MACUT \cite{wu2012memetic} &    TS-UBQP \cite{kochenberger2013solving} & TS/PM \cite{wang2013probabilistic} & MAMBP \cite{wu2013memetic} & MOH \\
\midrule
\endhead
\midrule
\endfoot

\hline \hline
\endlastfoot

        G1 &        800 &      11624 &      11624 &      11624 &      11624 &      11624 &      11624 &      11624 &      11624 \\

        G2 &        800 &      11620 &      11620 &      11620 &      11620 &      11620 &      11620 &      11617 &      11620 \\

        G3 &        800 &      11622 &      11622 &      11622 &      11622 &      11620 &      11620 &      11621 &      11622 \\

        G4 &        800 &      11646 &      11646 &      11646 &          - &      11646 &      11646 &      11646 &      11646 \\

        G5 &        800 &      11631 &      11631 &      11631 &          - &      11631 &      11631 &      11631 &      11631 \\

        G6 &        800 &       2178 &       2178 &       2178 &          - &       2178 &       2178 &       2177 &       2178 \\

        G7 &        800 &       2006 &       2006 &       2006 &          - &       2006 &       2006 &       2002 &       2006 \\

        G8 &        800 &       2005 &       2005 &       2005 &          - &       2005 &       2005 &       2004 &       2005 \\

        G9 &        800 &       2054 &       2054 &       2054 &          - &       2054 &       2054 &       2052 &       2054 \\

       G10 &        800 &       2000 &       2000 &       2000 &          - &       2000 &       2000 &       1998 &       2000 \\

       G11 &        800 &        564 &        564 &        564 &        564 &        564 &        564 &        564 &        564 \\

       G12 &        800 &        556 &        556 &        556 &        556 &        556 &        556 &        556 &        556 \\

       G13 &        800 &        582 &        582 &        582 &        582 &        580 &        582 &        582 &        582 \\

       G14 &        800 &       3064 &       3064 &       3064 &       3064 &       3061 &       3063 &       3062 &       3064 \\

       G15 &        800 &       3050 &       3050 &       3050 &       3050 &       3050 &       3050 &       3050 &       3050 \\

       G16 &        800 &       3052 &       3052 &       3052 &       3052 &       3052 &       3052 &       3052 &       3052 \\

       G17 &        800 &       3047 &       3047 &       3047 &          - &       3046 &       3047 &       3047 &       3047 \\

       G18 &        800 &        992 &        992 &        992 &          - &        991 &        992 &        992 &        992 \\

       G19 &        800 &        906 &        906 &        906 &          - &        904 &        906 &        905 &        906 \\

       G20 &        800 &        941 &        941 &        941 &          - &        941 &        941 &        941 &        941 \\

       G21 &        800 &        931 &        931 &        931 &          - &        930 &        931 &        930 &        931 \\

       G22 &       2000 &      13359 &      13359 &      13359 &      13359 &      13359 &      13349 &      13359 &      13359 \\

       G23 &       2000 &      13344 &      13342 &      13344 &      13344 &      13342 &      13332 &      13344 &      13344 \\

       G24 &       2000 &      13337 &      13337 &      13337 &      13337 &      13337 &      13324 &      13336 &      13337 \\

       G25 &       2000 &      13340 &      13340 &      13340 &          - &      13332 &      13326 &      13340 &      13340 \\

       G26 &       2000 &      13328 &      13328 &      13328 &          - &      13328 &      13313 &      13328 &      13328 \\

       G27 &       2000 &       3341 &       3341 &       3341 &          - &       3336 &       3325 &       3341 &       3341 \\

       G28 &       2000 &       3298 &       3298 &       3298 &          - &       3295 &       3287 &       3298 &       3298 \\

       G29 &       2000 &       3405 &       3405 &       3405 &          - &       3391 &       3394 &       3403 &       3405 \\

       G30 &       2000 &       3413 &       3413 &       3412 &          - &       3403 &       3402 &       3412 &       3413 \\

       G31 &       2000 &       3310 &       3310 &       3309 &          - &       3288 &       3299 &       3309 &       3310 \\

       G32 &       2000 &       1410 &       1410 &       1410 &       1410 &       1406 &       1406 &       1410 &       1410 \\

       G33 &       2000 &       1382 &       1382 &       1382 &       1382 &       1378 &       1374 &       1382 &       1382 \\

       G34 &       2000 &       1384 &       1384 &       1384 &       1384 &       1378 &       1376 &       1384 &       1384 \\

       G35 &       2000 &       7686 &       7686 &       7684 &       7686 &       7678 &       7661 &       7686 &       \textbf{7687} \\

       G36 &       2000 &       7680 &       7680 &       7678 &       7679 &       7670 &       7660 &       7678 &       7680 \\

       G37 &       2000 &       7691 &       7691 &       7689 &       7690 &       7682 &       7670 &       7689 &       7691 \\

       G38 &       2000 &       7688 &       7687 &       7687 &          - &       7683 &       7670 &       7688 &       7688 \\

       G39 &       2000 &       2408 &       2408 &       2408 &          - &       2397 &       2397 &       2408 &       2408 \\

       G40 &       2000 &       2400 &       2400 &       2400 &          - &       2390 &       2392 &       2400 &       2400 \\

       G41 &       2000 &       2405 &       2405 &       2405 &          - &       2400 &       2398 &       2405 &       2405 \\

       G42 &       2000 &       2481 &       2481 &       2481 &          - &       2469 &       2474 &       2481 &       2481 \\

       G43 &       1000 &       6660 &       6660 &       6660 &       6660 &       6660 &       6660 &       6659 &       6660 \\

       G44 &       1000 &       6650 &       6650 &       6650 &       6650 &       6639 &       6649 &       6650 &       6650 \\

       G45 &       1000 &       6654 &       6654 &       6654 &       6654 &       6652 &       6654 &       6654 &       6654 \\

       G46 &       1000 &       6649 &       6649 &       6649 &          - &       6649 &       6649 &       6649 &       6649 \\

       G47 &       1000 &       6657 &       6657 &       6657 &          - &       6656 &       6656 &       6657 &       6657 \\

       G48 &       3000 &       6000 &       6000 &       6000 &       6000 &       6000 &       6000 &       6000 &       6000 \\

       G49 &       3000 &       6000 &       6000 &       6000 &       6000 &       6000 &       6000 &       6000 &       6000 \\

       G50 &       3000 &       5880 &       5880 &       5880 &       5800 &       5880 &       5880 &       5880 &       5880 \\

       G51 &       1000 &       3848 &       3848 &       3848 &          - &       3847 &       3847 &       3847 &       3848 \\

       G52 &       1000 &       3851 &       3851 &       3851 &          - &       3849 &       3850 &       3851 &       3851 \\

       G53 &       1000 &       3850 &       3850 &       3850 &          - &       3848 &       3848 &       3850 &       3850 \\

       G54 &       1000 &       3852 &       3852 &       3852 &          - &       3851 &       3850 &       3851 &       3852 \\

       G55 &       5000 &      10299 &          - &      10294 &      10299 &      10236 &          - &      10299 &      10299 \\

       G56 &       5000 &       4017 &          - &       4012 &       4016 &       3934 &          - &       4016 &       \it{4016} \\

       G57 &       5000 &       3494 &          - &       3492 &          - &       3460 &          - &       3488 & 3494 \\

       G58 &       5000 &      19293 &          - &      19263 &          - &      19248 &          - &      19276 & \it{19288} \\

       G59 &       5000 &       6086 &          - &       6078 &          - &       6019 &          - &       6085 & {\bf 6087} \\

       G60 &       7000 &      14188 &          - &      14176 &      14186 &      14057 &          - &      14186 & {\bf 14190} \\

       G61 &       7000 &       5796 &          - &       5789 &          - &       5680 &          - &       5796 & {\bf 5798} \\

       G62 &       7000 &       4870 &          - &       4868 &          - &       4822 &          - &       4866 & \it{4868} \\

       G63 &       7000 &      27045 &          - &      26997 &          - &      26963 &          - &      26754 & \it{27033} \\

       G64 &       7000 &       8751 &          - &       8735 &          - &       8610 &          - &       8731 & \it{8747} \\

       G65 &       8000 &       5562 &          - &       5558 &       5550 &       5518 &          - &       5556 & \it{5560} \\

       G66 &       9000 &       6364 &          - &       6360 &       6352 &       6304 &          - &       6352 &      \it{6360} \\

       G67 &      10000 &       6950 &          - &       6940 &       6934 &       6894 &          - &       6934 &  \it{6942} \\

       G70 &      10000 &       9591 &          - &       9541 &          - &       9458 &          - &       9580 &       \it{9544} \\

       G72 &      10000 &       7006 &          - &       6998 &          - &       6922 &          - &       6990 &       \it{6998} \\

       G77 &      14000 &       9938 &          - &       9926 &          - &          - &          - &       9900 &  \it{9928} \\

       G81 &      20000 &      14048 &          - &      14030 &          - &          - &          - &      13978 & \it{14036} \\

 3dl101000 &       1000 &        896 &        896 &          - &          - &          - &          - &          - &        896 \\

 3dl102000 &       1000 &        900 &        900 &          - &          - &          - &          - &          - &        900 \\

 3dl103000 &       1000 &        892 &        892 &          - &          - &          - &          - &          - &        892 \\

 3dl104000 &       1000 &        898 &        898 &          - &          - &          - &          - &          - &        898 \\

 3dl105000 &       1000 &        886 &        886 &          - &          - &          - &          - &          - &        886 \\

 3dl106000 &       1000 &        888 &        888 &          - &          - &          - &          - &          - &        888 \\

 3dl107000 &       1000 &        900 &        900 &          - &          - &          - &          - &          - &        900 \\

 3dl108000 &       1000 &        882 &        882 &          - &          - &          - &          - &          - &        882 \\

 3dl109000 &       1000 &        902 &        902 &          - &          - &          - &          - &          - &        902 \\

3dl1010000 &       1000 &        894 &        894 &          - &          - &          - &          - &          - &        894 \\

 3dl141000 &       2744 &       2446 &       2446 &          - &          - &          - &          - &          - &       2446 \\

 3dl142000 &       2744 &       2458 &       2458 &          - &          - &          - &          - &          - &       2458 \\

 3dl143000 &       2744 &       2442 &       2442 &          - &          - &          - &          - &          - & {\bf 2444} \\

 3dl144000 &       2744 &       2450 &       2450 &          - &          - &          - &          - &          - &       2450 \\

 3dl145000 &       2744 &       2446 &       2446 &          - &          - &          - &          - &          - &       2446 \\

 3dl146000 &       2744 &       2452 &       2452 &          - &          - &          - &          - &          - &       2452 \\

 3dl147000 &       2744 &       2444 &       2444 &          - &          - &          - &          - &          - &       2444 \\

 3dl148000 &       2744 &       2448 &       2448 &          - &          - &          - &          - &          - &       2448 \\

 3dl149000 &       2744 &       2426 &       2426 &          - &          - &          - &          - &          - & {\bf 2428} \\

3dl1410000 &       2744 &       2458 &       2458 &          - &          - &          - &          - &          - &       2458 \\
\hline
    Better &            &   6/91/91 &    4/74/91 &   20/71/91 &    7/30/91 &   47/69/91 &   29/54/91 &   33/71/91 &            \\

     Equal &            &   73/91/91 &   70/74/91 &   51/71/91 &   23/30/91 &   22/69/91 &   25/54/91 &   37/71/91 &            \\

     Worse &            &    12/91/91 &    0/74/91 &    0/71/91 &    0/30/91 &    0/69/91 &    0/54/91 &    1/71/90 &            \\

\bottomrule
\end{longtable}
}
}

\section{Discussion}\label{sec_Discussion}
In this section, we investigate the role of several important ingredients of the proposed algorithm, including the descent improvement search operators $O_1$ and $O_2$ and the diversified improvement search operators $O_3$ and $O_4$. These studies are based on the same 10 challenging instances selected to determine the parameters (see Section \ref{subsec_para}). Only results for max-cut are presented in this section.

\subsection{Impact of the descent improvement search operators}\label{sec_ana_descent}
\begin{figure}
  \centering
  \subfigure[$f_{best-strategy} - f_{best known}$, gaps to best known objective values ]{
    \label{fig-dec-a} 
    \includegraphics[width=3.0in, height=2.30in]{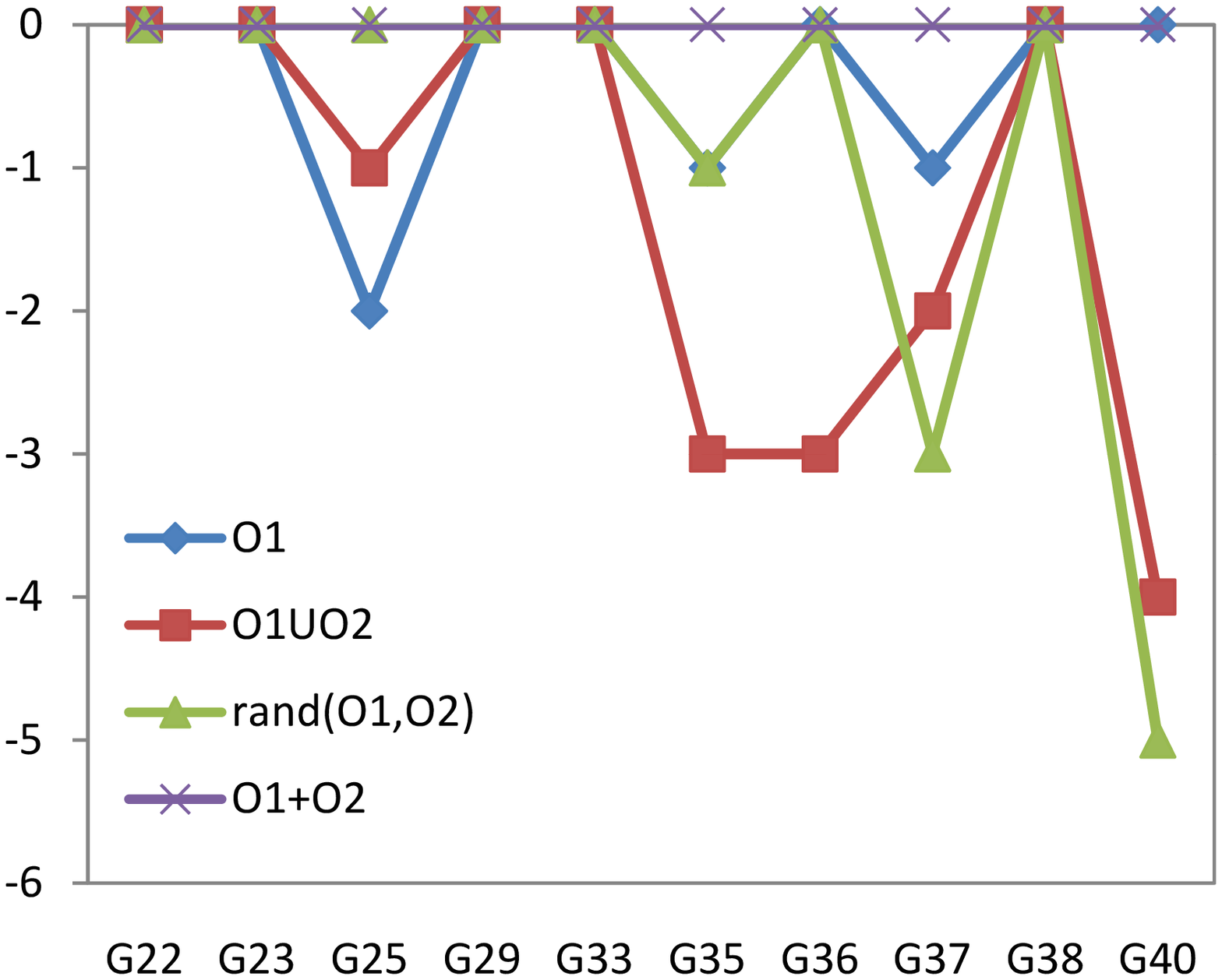}}
  \hspace{0.1in}
  \subfigure[$(f_{best known} - f_{avg-strategy}) / f_{best known} *100\%$, gaps to best known objective values]{
    \label{fig-dec-b} 
    \includegraphics[width=3.0in, height=2.3in]{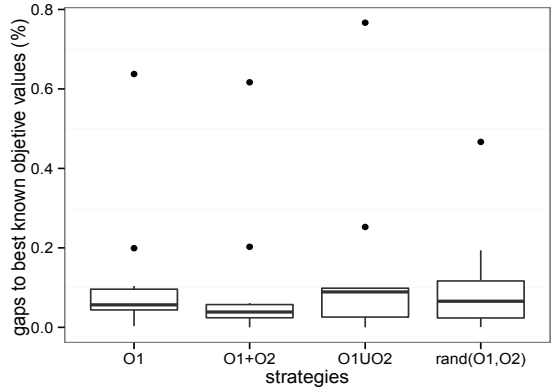}}
  \caption{Analysis of the move operators $O_1$, $O_2$}
  \label{fig-dec} 
\end{figure}

As described in Section \ref{subsec_descent_search}, the proposed algorithm employs operators $O_1$ and $O_2$ for its descent improvement phase to obtain local optima. To analyze the impact of these two operators, we implement three variants of our algorithm, the first one using the operator $O_1$ alone, the second one using the union $O_1\cup O_2$ such that the descent search procedure always chooses the best move among the $O_1$ \textit{and} $O_2$ moves \cite{lu2011neighborhood}, the third one using operator $rand(O_1,O_2)$ where the descent procedure applies randomly and with equal probability $O_1$ or $O_2$, while keeping all the other ingredients and parameters fixed as described in Section \ref{subsec_para}. The strategy used by our original algorithm is denoted as $O_1+O_2$, which is detailed in Section \ref{subsec_descent_search}. Each selected instance is solved 10 times by each of these variants and our original algorithm.  The stop criterion is a timeout limit of 30 minutes. The obtained results are presented in Table \ref{table_ana_dec}, including the best objective value $f_{best}$,  the average objective value $f_{avg}$ over the 10 independent runs, as well as the CPU times in seconds to reach $f_{best}$. To evaluate the performance, we calculate the gaps between the best objective values obtained by different strategies and the best objective values by our original algorithm, which is shown in Fig. \ref{fig-dec-a}. We also show in Fig. \ref{fig-dec-b} the box and whisker plots which indicates, for different $O_1$, $O_2$ combination strategies, the distribution and the ranges of the obtained results for the 10 tested instances. The results are expressed as the additive inverse of percent deviation of the averages results from the best known objective values obtained by our original algorithm.

From Fig. \ref{fig-dec-a}, one observes that for the tested instances, other combination strategies obtain fewer best known results compared to the strategy $O_1 + O_2$, and produce large gaps to the best known results on some instances.  From Fig. \ref{fig-dec-b}, we observe a clear difference in the distribution of the results with different strategies.  For the results with the strategies of $O_1 + O_2$, the plot indicates a smaller mean value and significantly smaller variation compared to the results obtained by other strategies.  We thus conclude that the strategy used by our algorithm ($O_1 + O_2$) performs better than other strategies.

\begin{table}\scriptsize
\centering
\caption{Comparative results for max-cut with varying combination strategies of $O_1$ and $O_2$ } \label{table_ana_dec}
\begin{tabular}{@{}lrrrrrrr@{}}
\toprule
Instance &   \multicolumn{ 3}{c}{$O_{1}$} &          & \multicolumn{ 3}{c}{$O_{1}\cup O_2$} \\
\cmidrule{2-4} \cmidrule{6-8}
         & $f_{best}$ & $f_{avg}$ & $time(s)$ &          & $f_{best}$ & $f_{avg}$ & $time(s)$ \\
\midrule
     G22 &    13359 &  13357.6 &   381.6 &          &    13359 &  13355.8 &  357.3 \\

     G23 &    13344 &  13343.6 &  473.4 &          &    13344 &    13344 &  550.9 \\

     G25 &    13338 &    13334 &   442.8 &          &    13339 &  13335.8 &  690.4 \\

     G29 &     3405 &  3398.22 &   211.1 &          &     3405 &   3396.4 &   254.2\\

     G33 &     1382 &   1381.4 &  553.5 &          &     1382 &     1382 &  716.5 \\

     G35 &     7686 &   7681.3 &  755.4 &          &     7684 &   7679.1 &  449.6 \\

     G36 &     7680 &     7672 &  1367.1 &          &     7677 &   7672.5 &  408.1 \\

     G37 &     7690 &   7685.5 &  1039.2 &          &     7689 &   7683.4 &  1099.0\\

     G38 &     7688 &     7684 &  135.2 &          &     7688 &   7681.2 &   177.8 \\

     G40 &     2400 &   2384.7 &  453.5 &          &     2396 &   2381.6 &  427.2 \\
\midrule
\toprule
Instance & \multicolumn{ 3}{c}{$rand(O_1,O_2)$} &          &  \multicolumn{ 3}{c}{$O_1+O2$} \\
\cmidrule{2-4} \cmidrule{6-8}
         & $f_{best}$ & $f_{avg}$ & $time(s)$ &          & $f_{best}$ & $f_{avg}$ & $time(s)$ \\
\midrule
     G22 &    13359 &    13356 &  365.3 &          &    13359 &    13357 &  438.2 \\

     G23 &    13344 &  13343.9 &  584.9 &          &    13344 &    13344 &  302.1 \\

     G25 &    13340 &  13336.4 &  408.8 &          &    13340 &  13335.5 &  451.5 \\

     G29 &     3405 &   3398.4 &  403.9 &          &     3405 &   3398.1 &    569.9 \\

     G33 &     1382 &   1381.8 &  585.2 &          &     1382 &   1381.4 &    667.4 \\

     G35 &     7686 &   7683.1 &  628.0 &          &     7687 &   7684.3 &  968.3 \\

     G36 &     7680 &     7672 &  944.8 &          &     7680 &   7675.3 &  1075.6 \\

     G37 &     7688 &   7681.7 &   1078.3 &          &     7691 &   7687.5 &   1133.2 \\

     G38 &     7688 &   7680.8 &   153.6 &          &     7688 &   7685.7 &  333.0 \\

     G40 &     2395 &   2388.8 &   412.4 &          &     2400 &   2385.2 &     467.1 \\
\hline
\end{tabular}

\end{table}

\subsection{Impact of the diversified improvement search operators} \label{sec_ana_div}
\begin{figure}
  \centering
  \subfigure[$f_{best-\rho} - f_{best known}$, gaps  between $f_{best}$ obtained with different $\rho$ values to best known objective values]{
    \label{fig-div-a} 
    \includegraphics[width=3.0in, height=2.3in]{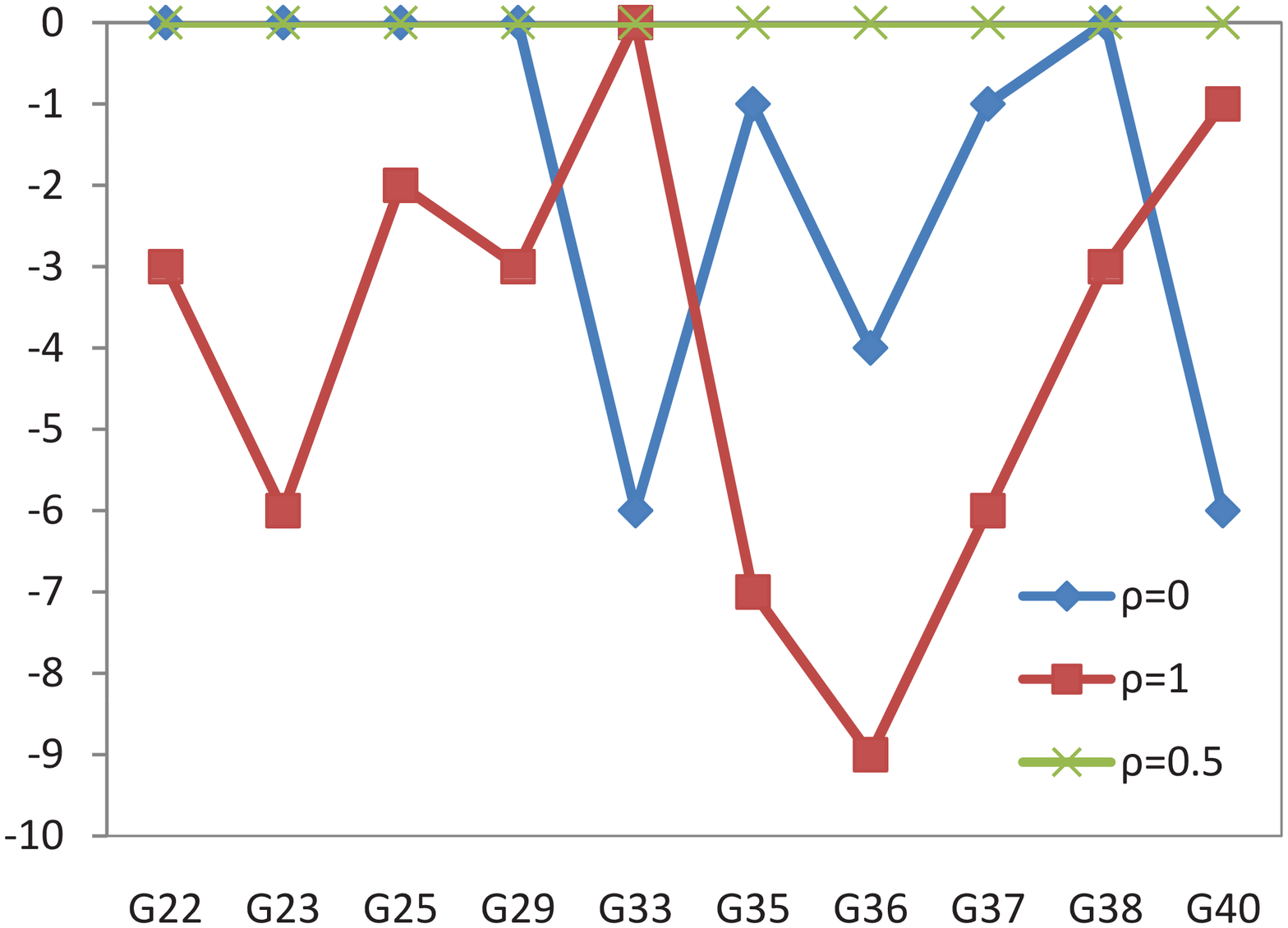}}
  \hspace{0.1in}
  \subfigure[$(f_{best known} - f_{avg-\rho}) / f_{best known} *100\%$,  gaps to best known objective values]{
    \label{fig-div-b} 
    \includegraphics[width=3.0in, height=2.30in]{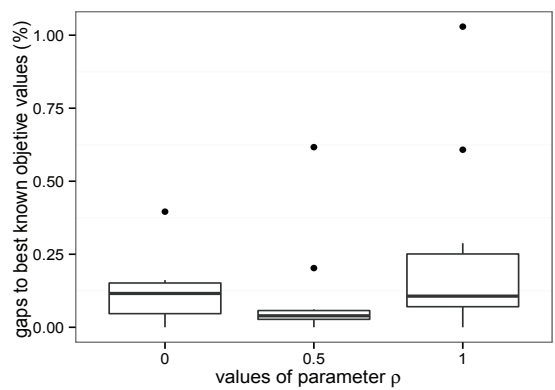}}
  \caption{Analysis of the move operators $O_3$, $O_4$}
  \label{fig-div} 
\end{figure}

As described in Section \ref{subsec_div_search}, the proposed algorithm employs two diversified operator $O_3$ and $O_4$ to enhance the search power of the algorithm and make it possible for the search to visit new promising regions. The diversified improvement procedure uses probability $\rho$ to select $O_3$ or $O_4$. To analyze the impact of operators $O_3$ and $O_4$, we test our algorithm with $\rho=1$ (using the operator $O_3$ alone), $\rho= 0.5$ (equal application of $O_3$ and $O_4$ used in our original MOH algorithm), $\rho=0$ (using the operator $O_4$ alone), while keeping all the other ingredients and parameters fixed as described before.  The stop criterion is a timeout limit of 30 minutes. We then independently solve each selected instance 10 times with those different values of $\rho$.  The obtained results are presented in Table \ref{table_ana_div}, including the best objective value $f_{best}$,  the average objective value $f_{avg}$ over the 10 independent runs, as well as the CPU times in seconds to reach $f_{best}$. To evaluate the performance, we again calculate the gaps between different best objective values shown in Fig. \ref{fig-div-a} and average objective values shown in Fig. \ref{fig-div-b}, where the set of values $f_{best}$, $f_{avg}$, when $\rho = 0.5$, are set as the reference values.

As Section \ref{sec_ana_descent}, to evaluate the performance, we calculate the gaps between the best objective values obtained with different values of $\rho$ and the best objective values by our original MOH algorithm ($\rho=0.5$), which is shown in Fig. \ref{fig-div-a}. We also show in Fig. \ref{fig-div-b} the box and whisker plots which indicates, for different values of $\rho$, the distribution and the ranges of the obtained results for the 10 tested instances. The results are expressed as the additive inverse of percent deviation of the averages results from the best known objective values obtained by our original algorithm.

Fig. \ref{fig-div-a} discloses that using $O_3$ or $O_4$ alone obtains fewer best known results than using them jointly and also achieves significantly worse results on some particular instances.  From Fig. \ref{fig-div-b}, we observes a visible difference in the distribution of the results with different strategies. For the results with the parameter $\rho = 0.5$,  the plot indicates a smaller mean value and significantly smaller variation compared to the results obtained by other strategies.  We thus conclude that jointly using $O_3$ and $O_4$ with $\rho = 0.5$ is the best choice since it produces better results in terms of both best results and average results.

\begin{table}\scriptsize
\caption{Comparative results for max-cut with varying parameter $\rho$} \label{table_ana_div}
\centering
\begin{tabular}{@{}lrrrrrrrrrrr@{}}
\toprule
    Instance &       \multicolumn{ 3}{c}{$\rho =1$} &            &       \multicolumn{ 3}{c}{$\rho =0$} &            &     \multicolumn{ 3}{c}{$\rho =0.5$} \\
\cmidrule{2-4} \cmidrule{6-8} \cmidrule{10-12}
    & $f_{best}$ &  $f_{avg}$ &  $time(s)$ &            & $f_{best}$ &  $f_{avg}$ &  $time(s)$ &            & $f_{best}$ &  $f_{avg}$ &  $time(s)$ \\
\midrule

     G22 &    13359 &  13350.1 &  352.7 &          &    13356 &  13355.2 &  440.6 &          &    13359 &    13357 &  438.2 \\

     G23 &    13344 &    13344 &  441.4 &          &    13338 &  13335.6 &   340.1 &          &    13344 &    13344 &  302.1 \\

     G25 &    13339 &  13335.1 &   426.1 &          &    13337 &  13333.5 &  412.9 &          &    13340 &  13335.5 &  451.5 \\

     G29 &     3405 &   3395.2 &   614.5 &          &     3402 &   3399.8 &   593.5 &          &     3405 &   3398.1 &    569.9 \\

     G33 &     1376 &   1373.6 &  519.9 &          &     1382 &     1382 &  609.2 &          &     1382 &   1381.4 &    667.7 \\

     G35 &     7686 &   7680.7 &   832.1 &          &     7680 &   7678.2 &   850.8 &          &     7687 &   7684.3 &  968.3 \\

     G36 &     7676 &   7669.2 &   1540.8 &          &     7671 &   7667.6 &  1304.8 &          &     7680 &   7675.3 &  1075.6 \\

     G37 &     7690 &   7681.2 &   1167.8 &          &     7685 &   7679.6 &   1053.8 &          &     7691 &   7687.5 &   1133.2 \\

     G38 &     7688 &   7681.4 &  275.1 &          &     7685 &     7679 &  257.3 &          &     7688 &   7685.7 &  333.0 \\

     G40 &     2394 &   2375.3 &  453.0 &          &     2399 &   2390.5 &   529.8 &          &     2400 &   2385.2 &     467.1 \\
\hline

\end{tabular}
\end{table}

\section{Conclusion}\label{sec_Conclusion}
Our multiple search operator algorithm (MOH) for the general max-k-cut problem achieves a high level performance by including five distinct search operators which are applied in three search phases. The descent-based improvement phase aims to discover local optima of increasing quality with two intensification-oriented operators. The diversified improvement phase combines two other operators to escape local optima and discover promising new search regions. The perturbation phase is applied as a means of strong diversification to get out of deep local optimum traps. To obtain an efficient implementation of the proposed algorithm, we developed streamlining techniques based on bucket structures.

We demonstrated the effectiveness of the MOH algorithm both in terms of solution quality and computation efficiency by a computational study on the two sets of well-known benchmarks composed of 91 instances. For the general max-k-cut problem, the proposed algorithm is able to improve 90 percent of the current best known results available in the literature. Moreover, for the very popular special case with $k=2$, i.e., the max-cut problem, MOH also performs extremely well by improving 6 best known results which were previously established by any max-cut algorithms of the literature including several recent algorithms published since 2012.

We also investigated alternative strategies for combing search operators and justified the combination adopted in the proposed MOH algorithm.

Given that most ideas of the proposed algorithm are general enough, it is expected that they can be useful to design effective heuristics for other graph partitioning problems.

\section*{Acknowledgment}
The work is partially supported by the LigeRo project (2009-2014) from the Region of Pays de la Loire (France) and the PGMO (2014-0024H) project from the Jacques Hadamard Mathematical Foundation. Support for Fuda Ma from the China Scholarship Council is also acknowledged.




\end{document}